\documentclass[usenatbib]{mnras}
\usepackage[T1]{fontenc}
\usepackage{ae,aecompl}
\usepackage{graphicx}
\usepackage{amsmath}
\usepackage{amssymb}	
\usepackage{subfig}
\usepackage{multirow}
\usepackage{tabularx}
\usepackage{pifont}
\usepackage{array}
\usepackage{xcolor}
\usepackage{enumerate}
\usepackage{amsfonts}
\usepackage{scalefnt}
\usepackage[normalem]{ulem}
\usepackage{tikz}
\usepackage{float}
 
\usepackage{booktabs} % For table lines
\newcolumntype{P}[1]{>{\centering\arraybackslash}p{#1}}
\def\simless{\mathbin{\lower 3pt\hbox
{$\rlap{\raise 5pt\hbox{$\char'074$}}\mathchar"7218$}}}   %< or of order
\def\simmore{\mathbin{\lower 3pt\hbox
{$\rlap{\raise 5pt\hbox{$\char'076$}}\mathchar"7218$}}}   %> or of order
\newcommand{\be}{\begin{equation}}
\newcommand{\ee}{\end{equation}}
\usepackage[section]{placeins}
\newcommand{\MSun}{{\rm M}_{\sun}}
\newcommand{\mach}{\mathcal{M}_{\rm{a}}}
\newcommand{\machz}{\mathcal{M}_{\rm{a0}}}

\title[The imprint of gas on GWs from LISA IMBHBs]{The imprint of gas on gravitational waves from LISA intermediate-mass black hole binaries}
\author[M. Garg et al.]{Mudit Garg,\thanks{E-mail: mudit.garg@ics.uzh.ch} Andrea Derdzinski, Lorenz Zwick, Pedro~R. Capelo and Lucio Mayer\\
Center for Theoretical Astrophysics and Cosmology, Institute for Computational Science, University of Zurich,\\
Winterthurerstrasse 190, CH-8057 Z\"urich, Switzerland}  

\date{Received / Accepted}
\pubyear{2022}
\begin{document}
\label{firstpage}
\pagerange{\pageref{firstpage}--\pageref{lastpage}}
\maketitle

\begin{abstract}
We study the effect of torques on circular inspirals of intermediate-mass black hole binaries (IMBHBs) embedded in gas discs, wherein both BH masses are in the range $10^2$--$10^5~\MSun$, up to redshift $z = 10$.  We focus on how torques impact the detected gravitational wave (GW) waveform in the LISA frequency band when the binary separation is within a few hundred Schwarzschild radii. For a sub-Eddington accretion disc with a viscosity coefficient $\alpha=0.01$, surface density $\Sigma\approx10^5$~g~cm$^{-2}$, and Mach number $\mach\approx80$, a gap, or a cavity, opens when the binary is in the LISA band. Depending on the torque's strength, LISA will observe dephasing in the IMBHB's GW signal up to either $z\sim5$ for high mass ratios ($q\approx0.1$) or to $z\sim7$ for $q\approx10^{-3}$. We study the dependence of the measurable dephasing on variations of BH masses, redshift, and accretion rates. Our results suggest that phase shift is detectable even in high-redshift ($z = 10$) binaries if they experience super-Eddington accretion episodes. We investigate if the disc-driven torques can result in an observable `time-dependent' chirp mass with a simplified Fisher formalism, finding that, at the expected signal-to-noise ratio, the gas-induced variation of the chirp mass is too small to be detected. This work shows how gas-induced perturbations of vacuum waveforms should be strong enough to be detected by LISA for the IMBHB in the early inspiral phase. These perturbations encode precious information on accretion discs and galactic nuclei astrophysics. High-accuracy waveform models which incorporate these effects will be needed to extract such information.
\end{abstract} 


\section{Introduction}\label{Sec1}

In the 2030s, the Laser Interferometer Space Antenna (LISA; \citealt{AmaroSeoane2017,Barack2019}) will be launched into space and will follow Earth around its orbit of the Sun. It will be sensitive to gravitational waves (GWs) within the observed frequency range of $\sim$10$^{-4}$--$10^{-1}$~Hz. Primary source targets of LISA include the coalescence of black hole binaries (BHBs) with primary masses $10^6$--$10^{8}~\MSun$, which are called supermassive BHs (SMBHs), as well as inspirals of BHBs with secondary-to-primary mass ratio $q = 10^{-6}$--$10^{-3}$, termed extreme or intermediate mass-ratio inspirals (E/IMRIs; \citealt{Babak2017,AmaroSeoane2018}). Another important source for LISA will be intermediate-mass BHBs (IMBHBs), wherein both companions' masses are in the range $10^2$--$10^5~\MSun$ \citep{Miller2004}.

The detection of even a single IMBH has been elusive in the past. Currently, we have hundreds of IMBH candidates in our local Universe (see, e.g. \citealt{Mezcua2017,Greene2020} for a review), and recently, we have detected three well-constrained IMBHs. The first one, detected electromagnetically from a survey of active dwarf galaxies, which may host an IMBH at their centres, is a $\sim$50000~$\MSun$ BH termed RGG 118 at $z=0.0243$ \citep{Baldassare2015}. The second one, detected either in a large star cluster or a tidally stripped dwarf galaxy via X-ray continuum fitting, is a $\sim$20000~$\MSun$ BH termed TDE J2150 \citep{Lin2018,Wen2021}. And the last one, detected via GWs, is the $\sim$150~$\MSun$ BH remnant of the BHB merger event GW190521 at $z=0.82$ \citep{Abbott2020}. However, we only have a single contender (CR7 at $z=6.6$; \citealt{Hartwig2016}) for IMBHs at high redshift ($z\gtrsim4$) and no candidate for IMBHB in our Universe. We may find new candidates for IMBHs and IMBHBs with next-generation electromagnetic (EM) surveys or GW detectors, e.g. with the James Webb Space Telescope (JWST; \citealt{Natarajan2017,Cann2018}), the square kilometre array (SKA; \citealt{Whalen2021}), LISA, TianQin \citep{Wang2019}, and TAIJI \citep{Gong2021}, among others. LISA's sensitivity would allow us to obtain the first unambiguous evidence of the existence of IMBHBs, even up to redshift $z\sim20$. 

One of the most promising channels for forming tight IMBHBs (sub-pc separation) is through the interaction of seed BHs. There is significant interest in the seed growth mechanisms in the literature due to observations of billion solar mass SMBHs in the form of quasars as early as $z\gtrsim7$ \citep[see, e.g.][]{Banados2017,Matsuoka2019,Yang2020}. How these BH seeds form and their rapid journey to become SMBHs is an ongoing debate in the literature. The two prominent candidates for seeding models are light seeds and heavy seeds (see, e.g. \citealt{Valiante2017,Woods2019} for a review). Light seeds are remnants of metal-free Pop-III stars at $z\gtrsim20$ with masses $\sim$10--100~$\MSun$. These seeds can subsequently accrete in the presence of large amounts of gas, sometimes at super-Eddington rates, which could lead to the formation of SMBHs at $z\sim7$ \citep{Natarajan2014,Volonteri2010}. However, maintaining super-Eddington accretion rates for significant periods is challenging. High-redshift galaxies tend to be clumpy, meaning that BHs only seldom intersect with a gas reservoir \citep{Smith2018}. Furthermore, feedback from accretion can quickly evacuate the typical surroundings of an IMBH \citep{Regan2019}. Also, accretion flows can self-regulate by changing their radiative efficiency \citep{Lupi2016,Sassano2022}. Heavy seeds originate from the direct collapse BH (DCBH) scenario, wherein pristine or low-metallicity gas halos \citep{Loeb1994} or supermassive stars \citep{Haemmerle2020} directly collapse into a $\sim$10$^4$--10$^6~\MSun$ mass BH. Alternatively, more massive seeds can originate from a rapid assembly of halos \citep{Wise2019} or a merger of gas-rich galaxies \citep{Mayer2010,Mayer2015,Mayer2017,Mayer2019}. 

There are several other formation channels of tight IMBHBs. They could form during coalescing active dwarf galaxies. One other possibility for merging IMBHBs may arise in the active galactic nuclei (AGN) discs of massive galaxies (see, e.g. \citealt{Tagawa2020}), where in-situ star formation or orbit capture of nearby stars and BHs can lead to hierarchical BH mergers (see, e.g. \citealt{McKernan2012}), the building blocks of which fall into the LIGO-Virgo-KAGRA band (e.g. the $\sim$150~$\MSun$ GW190521 event; \citealt{Abbott2020}). Diverse growth and binary formation mechanisms lead to different predictions for detection rates of coalescing IMBHBs by LISA. Rates depend on various uncertainties in seed formation and binary interactions, e.g. the efficiency of metal enrichment of heavy seeds due to star formation feedback \citep{Sesana2007}, delay between galaxy merger and BHB coalescence (which is mediated by dynamical friction, stellar hardening, gas or third-body interaction), or supernova feedback \citep{Bonetti2018,Barausse2020}. These models predict the detection of $\sim$2--100~[yr$^{-1}$] IMBHB coalescence by LISA. 

We can reasonably expect a gas-rich environment around IMBHs for $z\gtrsim1$. Active dwarf galaxies tend to be gas rich, and we have several candidates for them in the local Universe \citep{Mezcua2017,Greene2020}. We expect gaseous accretion discs to be able to form around nascent SMBHs below $z\sim10$ (see, e.g. \citealt{Lodato2006,Regan2009,Pfister2019}), which could also be the case for $\sim$10$^{4-5}~\MSun$ IMBHs. Additionally, dark matter (DM) halos could merge and enhance gas-inflow rates due to tidal torques and shocks, justifying the presence of gas around multiple BHs \citep{Wise2019}. 

If gas is present, it can influence the orbital evolution of the binary through dynamical friction or torques. In some cases, particularly for small component masses, dephasing of the inspiral due to gas can produce detectable signatures in the GW waveform (see, e.g. \citealt{Kocsis2011,Yunes2011,Barausse2014,Derdzinski2019,Derdzinski2021,Zwick2021,Zwick2022} for E/IMRIs). However, the gas imprints are potentially detectable for only a subset of E/IMRIs at $z\lesssim1$. \citet{Barausse2014} explored dephasing for EMRIs and near-equal mass BH binaries embedded in a thin accretion disc assuming a one-year LISA observation run and found that we could detect a phase shift due to dynamical friction. However, their work neglected nuances of binary-disc interaction, which could result in a gap/cavity formation and torques due to gas flowing into a gap/cavity created by BHBs \citep{MacFadyen2008,Haiman2009,Duffell2014,Farris2014,DOrazio2015,Franchini2022}.

This work studies the effect of gas torques on IMBHBs embedded in a thin gaseous disc, assuming a four-year observation run for LISA. We are interested in this particular system because, if they exist, IMBHBs will spend up to $\sim$10~years in the LISA band with the initial binary rest-frame separation up to a few hundred Schwarzschild radii; therefore, they are the `golden' sources for LISA \citep[e.g.][]{AmaroSeoane2022}. Furthermore, IMBHB's small companion masses suggest that, similar to E/IMRIs, gas effects may impart a small but measurable contribution to the GW evolution. We only study gas torques in line with similar studies for EMRIs in the LISA band \citep{Yunes2011,Derdzinski2019,Derdzinski2021}. We also explore whether the smaller companion (or binary) would open a gap (or cavity) and how that would affect the strength of the gas torque.

LISA will be able to detect massive BHBs with high signal-to-noise ratios \citep[SNRs;][]{AmaroSeoane2017} and this should allow us to run highly accurate parameter estimation pipelines to recover binary parameters and potentially learn about the environment in which these systems are evolving \citep[see, e.g.][]{Cardoso2020}. The detection of environmental signatures in GWs will allow us to better resolve the presence of gas around BH binaries in regions that are currently electromagnetically unresolvable, which could constrain the growth of IMBHs, at least those that merge with another. Moreover, such measurements could complement the detection of EM counterparts (see, e.g. \citealt{Baker2019}), which should also occur when gas is present and can help with the binary localisation. For example, an `EM chirp' is caused by the production of a characteristic X-ray light curve, which can occur either due to the variable accretion in the inspiral \citep{Farris2015,Tang2018} or due to relativistic Doppler modulations and lensing effects \citep{Haiman2017}. There will be exciting synergies of LISA with the next-generation X-ray detector Athena \citep{Piro2022} and the radio detector SKA \citep{Tamanini2016}. Furthermore, gas imprints on GW waveforms could constrain gas torques on the binary. These torques have a nonlinear dependence on disc parameters, e.g. studies predict both inward and outward torques \citep{Duffell2020,Munoz2020,Tiede2020,Derdzinski2019,Derdzinski2021}. The presence of gas could also induce biases and degeneracies in the binary parameter estimation from LISA data, which we address. 

The paper is structured as follows. In Section~\ref{Sec2}, we provide the basics of GWs and discuss IMBHBs, which will be detectable by LISA. Section~\ref{Sec3} describes the accretion disc model we use and the  assumptions we have on our fiducial binary. Section~\ref{Sec4} studies the migration of IMBHBs under the influence of gas torques in the regime wherein a gap or cavity is carved in the disc. In Section~\ref{Sec5}, we compute the dephasing on the GW signal induced by gas and we show how this scales with the accretion rate on to the binary. We also explore the effect of gas on parameter estimation, and we determine for which critical accretion disc parameters the dephasing is detectable. We discuss the findings in Section~\ref{Sec6} and summarize the key takeaways of this work in Section~\ref{Sec7}. 

\section{A primer on GW and IMBHBs in the LISA band}\label{Sec2}

In this section, we lay out the essential framework and terminology necessary for our analysis. Let us first consider two non-spinning BHs in vacuum, with a rest-frame primary mass $M_{\rm{p}}$ and secondary mass $M_{\rm{s}}\leq M_{\rm{p}}$ (with $M_{\rm tot}\equiv M_{\rm p}+M_{\rm s}$), revolving around each other in circular orbits with a rest-frame separation $r$. Their rest-frame orbital angular frequency is $\Omega\equiv\sqrt{GM_\text{tot}/r^3}$, where $G$ is the gravitational constant. GWs radiated due to their interaction have a rest-frame frequency $f_{\rm r}$, which is two times the orbital frequency:

\begin{equation}\label{Eq1}
    {f_{\rm{r}}}=\frac{1}{\pi}\left(\frac{GM_\text{tot}}{r^3}\right)^\frac12.
\end{equation}

At Earth, ${f_{\rm{r}}}$ is redshifted to an observed frequency $f$ via the relation ${f_{\rm{r}}}={f}(1+z)$. If circular inspirals of BHBs are driven solely by GWs, then their orbital evolution rate is given by \citep{Peters1964}

\begin{equation}\label{Eq2}
    \dot{r}_\text{\scriptsize{GW}}=-\frac{64}{5}\frac{G^3}{c^5}\frac{qM_{\rm{p}}^2M_\text{tot}}{r^3},
\end{equation}

\noindent where $q\equiv M_{\rm{s}}/M_{\rm{p}}\leq1$ is the mass ratio and $c$ is the speed of light in vacuum. We are assuming zero eccentricity for this work. We neglect higher-order post-Newtonian (PN) terms for the present work because, as we will show later, gas signatures on the GW waveform are more prominent when the binary separation is more than a few tens of Schwarzschild radii. Using this rate, we can calculate the total number of orbits spent by a binary between two separations, $r_{\rm{max}}$ and $r_{\rm{min}}$, as

\begin{equation}\label{Eq3}
    N_\text{orb}=\frac{1}{2}\int^{r_{\rm{min}}}_{r_{\rm{max}}} {\rm d}r\frac{{f_{\rm{r}}}}{\dot{r}_\text{\scriptsize{GW}}}.
\end{equation}

The GW strain amplitude $h$ depends upon the binary's redshift and chirp mass $\mathcal{M}\equiv(M_{\rm{p}}^2 q)^{3/5}/M_\text{tot}^{1/5}$. The dependence on the chirp mass is due to Eq.~\eqref{Eq2}, and $\mathcal{M}$ is also the best measured binary parameter (see, e.g. \citealt{Cutler1994}). We can also express $\mathcal{M}$ in terms of ${f_{\rm{r}}}$ and its rest-frame time derivative $\dot{f}_{\rm{r}}$ \citep{Abbott2016}:

\begin{equation}\label{Eq4}
    \mathcal{M}=\frac{c^3}{G}\left(\frac{5}{96}\pi^{-\frac83}{f}^{-\frac{11}{3}}_{\rm{r}}\dot{f}_{\rm{r}}\right)^{\frac35}.
\end{equation}

The sky- and polarization-averaged strain $h$ of a source at the comoving coordinate distance $r(z)$ (see, e.g. \citealt{Ryden2016}) is given by \citep{Sesana2005}

\begin{equation}\label{Eq5}
    h = \frac{8\pi^\frac23}{10^\frac12}\frac{(G\mathcal{M})^\frac53}{c^4}\frac{{f_{\rm{r}}}^\frac23}{r(z)}.
\end{equation}

We analyze GWs until the separation reaches the innermost stable circular orbit (ISCO) of the primary BH, i.e. $r_\text{ISCO}\equiv 3r_{\rm{s}}$, where $r_{\rm{s}} \equiv2GM_{\rm{p}}/c^2$ is the Schwarzschild radius of the primary BH (or until the binary leaves the LISA band, if $r_{\rm ISCO}$ has not been reached yet). Another important quantity is the total number of cycles spent by a source at a given observed frequency ${f}$ \citep{Sesana2005}:

\begin{equation}\label{Eq6}
    n\approx \frac{f^2}{\dot{f}}.
\end{equation}

In practice, we can observe the true GW cycles only for  ${f}>{f}_\text{knee}\equiv n/\tau$, where $\tau$ is the mission lifetime. Otherwise, binary spans maximum $n={f}\tau$ orbits.

We can only observe IMBHBs whose GWs characteristic strain $h_{\rm{c}}$ is above the LISA sensitivity curve in the observed GW frequency domain. The characteristic strain is a visualization aid, which modifies the observed GW amplitude by taking into account how signal adds up as a source spends more time in the detector \citep{Moore2014}. Effectively, the SNR of a given source is given by the area between $h_{\rm{c}}$ of a source and the detector sensitivity. For this work, we assume that the LISA arm length is $2.5 \times 10^{11}$~cm, and its mission lifetime $\tau=4$ years, and use the sensitivity curve presented in \citet{RobsonCornishChang2019}.

Now we have all the tools to define the characteristic strain

\begin{equation}\label{Eq7}
    h_{\rm{c}}=\begin{cases} 
        h\sqrt{{f}\tau}, & {f} \leq {f}_\text{knee}, \\
        h\sqrt{n}, & {f}_\text{knee}<{f} \leq {f}_\text{ISCO}, \\
        0, & {f}_\text{ISCO}<{f},
   \end{cases}
\end{equation}

\noindent where ${f}_\text{ISCO}$ is the observed frequency when the secondary BH is at the primary BH's ISCO. If we hope to distinguish environmental effects on the GW signal, then we need to have a sufficient SNR, which can be computed between two observed frequencies ${f}_{\text{min}}$ and ${f}_{\text{max}}$ as

\begin{equation}\label{Eq8}
    \text{SNR}=\sqrt{2 \cdot 4\int^{{f}_{\text{max}}}_{{f}_{\text{min}}}{\rm d}{f}^\prime\frac{h_{\rm{c}}^2({f}^\prime)}{S_{\rm{t}}({f}^\prime){f}^{\prime2}}},
\end{equation}

\noindent where $S_{\rm{t}}({f})$ (with units Hz$^{-1}$) is the LISA sensitivity profile given by \citet{RobsonCornishChang2019}. The pre-factors 2 and 4 come from currently planned six links for LISA and one-sided spectral noise density normalization, respectively. As customary, we adopt a detectability threshold of SNR $\geq8$ (same as in \citealt{Bonetti2018,Barausse2020}).

Our fiducial binary consists of two IMBHs with primary masses in the range $10^2$--$10^5~\MSun$ and mass ratios between $10^{-3}$--$10^{-1}$. This is motivated by results of the Millennium simulation suite \citep{Fakhouri2010}, which finds that unequal-mass ratio mergers are more likely than equal-mass ones. We focus on results for $q=0.1$ as a reference because the corresponding binaries will have higher SNR. 

\begin{figure}
    \centering
    \includegraphics[width=1\linewidth]{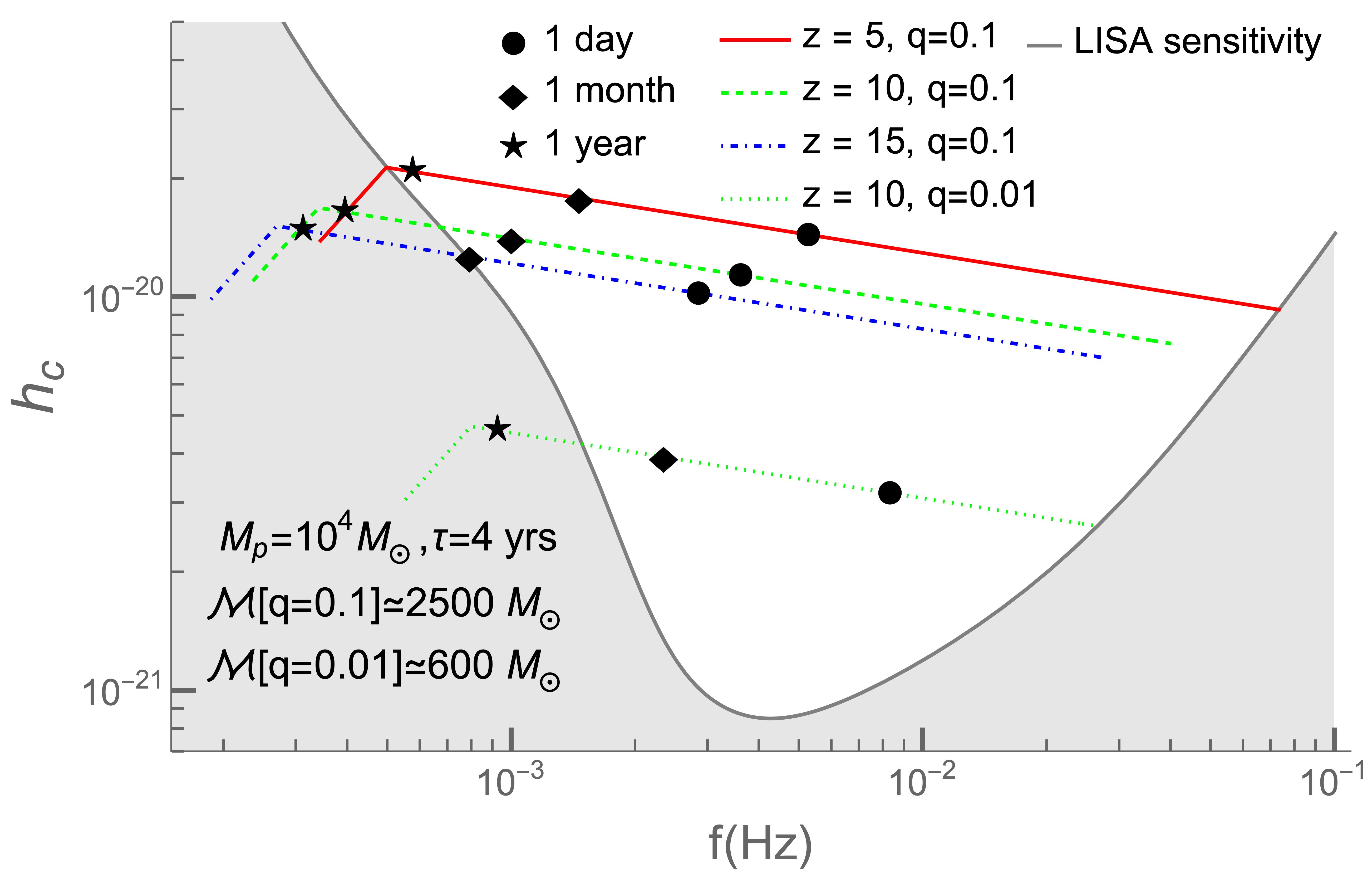}
    \caption{Characteristic strain $h_{\rm{c}}$ for IMBHBs with $M_{\rm{p}}=10^4~\MSun$ and $q=0.1$ for redshifts $z=5$ (solid red), 10 (dashed green), and 15 (dot-dashed blue), during the last four years before reaching $r_{\rm ISCO}$. For comparison, we have included $h_{\rm{c}}$ for $M_{\rm{p}}=10^4~\MSun,~q=0.01,$ and $z=10$ (dotted green). We have also marked when these binaries are 1 day ($\bullet$), 1 month ({\tiny $\blacklozenge$}), and 1 year ({\scriptsize $\bigstar$}) away from $r_{\rm ISCO}$. The grey curve marks the LISA sensitivity curve.}
    \label{Fig1}
\end{figure}

In Fig.~\ref{Fig1}, we show $h_{\rm{c}}$ for IMBHBs with $M_{\rm{p}}=10^4~\MSun$ and $q=0.1$, in the four-year observation window of LISA for three redshifts, $z=~5,~10,~\text{and}~15$, until their respective ISCO frequencies. We also mark when binaries are one day, one month, and one year away from reaching $r_{\rm ISCO}$. There is more frequency chirp towards the end of their evolution because of the stronger radiation of energy. Furthermore, higher-redshift binaries spend less time in the band with relatively lower $h_{\rm{c}}$ and this translates to lower $N_\text{orb}$ as well as SNR. We also show the case $M_{\rm{p}}=10^4~\MSun,~q=0.01,$ and $z=10$, which has weaker strain, yet still attains substantial SNR.

\begin{figure}
\centering
\subfloat[]{\includegraphics[width=1\linewidth]{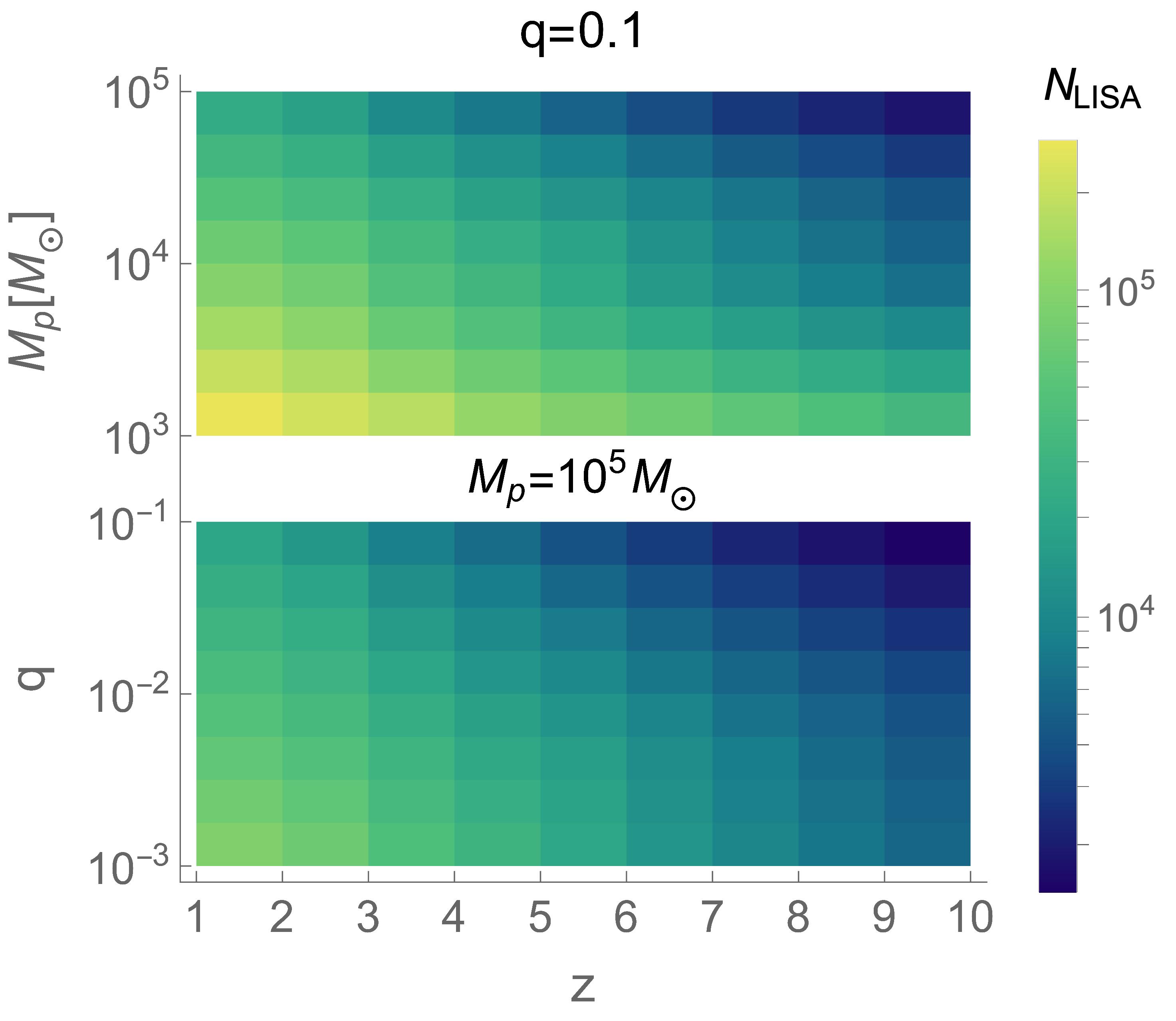}}\\
\subfloat[]{\includegraphics[width=1\linewidth]{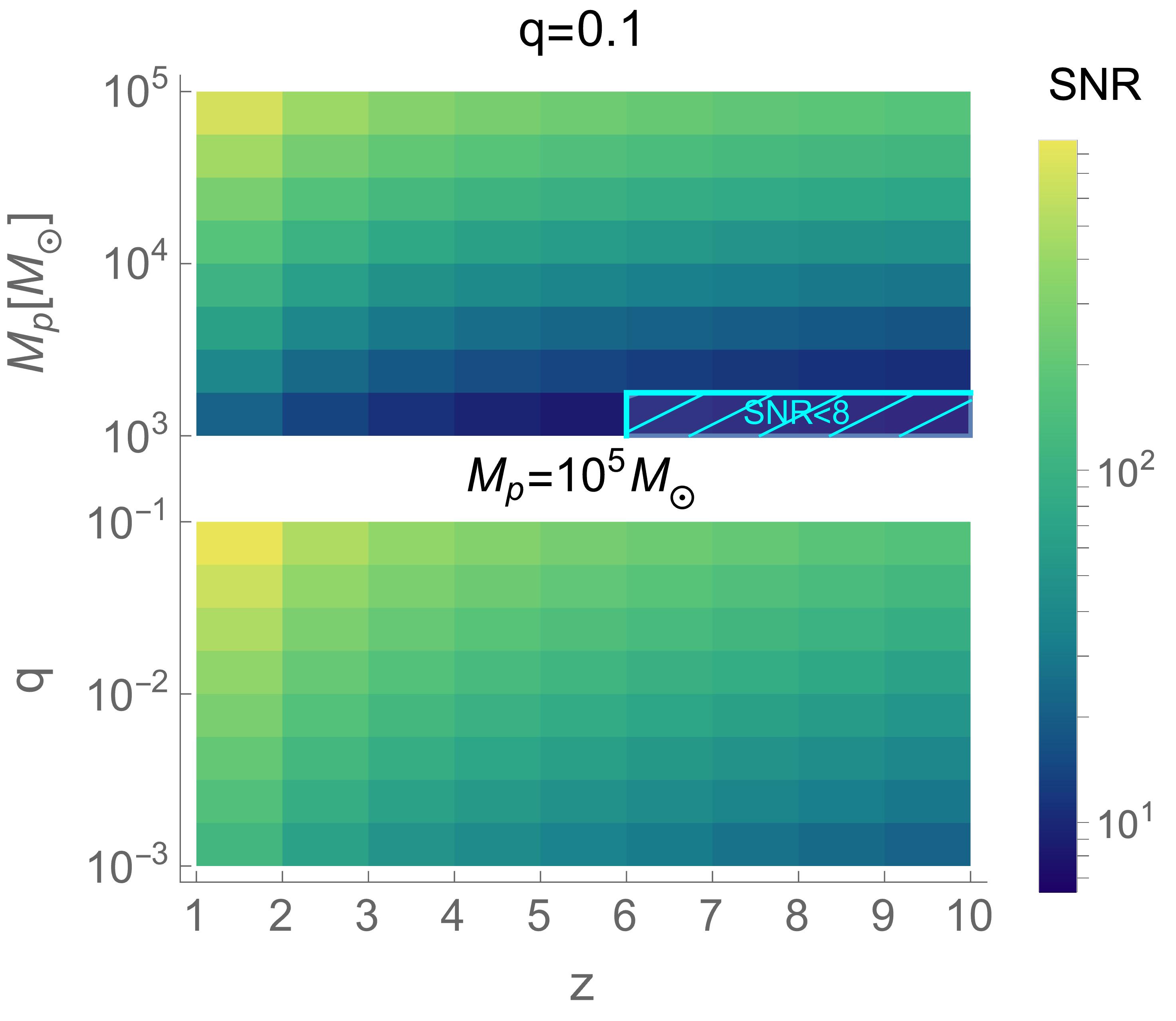}}
\caption{(a) Total number of orbits and (b) accumulated SNR in the LISA band for two different parameter combinations relevant for IMBHBs. In each pair of plots, in the top panel, we fix $q=0.1$ and vary $M_{\rm{p}}$ from $10^3$ to $10^5~\MSun$ and $z$ from 1 to 10, and in the bottom panel, we keep the primary mass $M_{\rm{p}}=10^5~\MSun$ constant and increase $q$ from $10^{-3}$ to $10^{-1}$ and $z$ from 1 to 10. The bottom panel has a LISA non-detectability region (SNR$<8$) shaded with diagonal cyan lines.}
\label{Fig2} 
\end{figure}

In Fig.~\ref{Fig2}, we compute the total number of orbits $N_\text{LISA}$ and the SNR in the LISA observation window for different systems relevant to IMBHBs.\footnote{Our SNR estimates are consistent with \citealt{AmaroSeoane2017}.} We assume $r_{\rm{min}}$ ($f_{\rm{max}}$) in Eq.~\eqref{Eq3} (Eq.~\ref{Eq8}) to be $r_{\rm{f}}$ ($f_{\rm{f}}$), which is the separation (observed frequency) when a binary leaves the LISA band or, if the separation reaches $r_{\rm ISCO}$ before then, is $r_{\rm{ISCO}}$ ($f_{\rm{ISCO}}$) and $r_{\rm{max}}$ ($f_{\rm{min}}$) to be $r_{\rm{i}}$ ($f_{\rm{i}}$), which is the minimum separation (maximum observed frequency) between when a binary enters the LISA band, or when the binary is traced back four years from $r_{\rm{f}}$ ($f_{\rm{f}}$). The first (second) pair of plots shows $N_{\rm LISA}$ (SNR) as a function of redshift and primary mass (for a fixed mass ratio; top panel) or mass ratio (for a fixed primary mass; bottom panel). We only show these quantities for redshifts 1--10, although binaries may merge at $z > 10$ (especially for seed BHs). However, as we will show in Section~\ref{Sec5}, this redshift range is the most relevant for detecting possible environmental effects. 

For these parameters, IMBHBs spent at most $\sim10$~years in the LISA band, which we computed by integrating Eq.~\eqref{Eq2}. Moreover, for the given redshift range, binaries with $M_{\rm{p}}\lesssim10^{3.75}~\MSun$ and $q=0.1$ merge outside the band yet still attain significant SNR in the LISA observation window, which allows us to analyze their properties. Almost all IMBHBs for the range of given parameters are detectable except for primary masses below $\sim10^{3.25}~\MSun$, mass ratios $\sim$0.1, and redshifts above $\sim$6.

In the next section, we discuss the accretion disc solutions we use to study the effects of gas on BH binaries.

\section{Accretion disc model}\label{Sec3}

We explore the effects of gas on the dynamics of our systems of interest. This is a relevant study for IMBHs embedded in gas-rich galactic nuclei, which is likely to be the case for primary MBHs in the mass range accessible to LISA and even more so at high redshift, where galaxies are increasingly gas-rich.

To this aim we adopt the thin accretion disc model detailed in Appendix \ref{AppA}, which follows the canonical model from \citet{ShakuraSunyaev1973}. The disc is radiatively efficient, optically thick, and geometrically thin, such that the aspect ratio $h/r<1$,  where $h$ is the disc height and $r$ is the distance to the central BH of mass $M$. Assuming a steady-state solution, the disc has a constant mass accretion rate $\dot{M}=3\pi\nu\Sigma$, where $\Sigma(r)$ is the surface density profile and $\nu$ is the kinematic viscosity, for which we assume a turbulent viscosity prescription: $\nu\equiv\alpha c_{\rm{s}}h$, where $\alpha<1$ is the viscosity coefficient and $c_{\rm{s}}$ is the speed of sound. The disc has a Mach number profile $\mach=v_\phi/c_{\rm{s}}= r/h$, where $v_\phi= \sqrt{GM/r}$ is the azimuthal velocity. We compute solutions of the disc model with three different single IMBHs of masses $M= 10^3$, $10^4$, and $10^5~\MSun$ embedded at the centre of the disc and produce profiles for the disc surface density $\Sigma$ and the Mach number $\mach$ over the range of 10--1000 $r_{\rm{s}}$ of the central body for $\alpha=0.01$ and $\dot{M}=0.1\dot{M}_\text{Edd}$, where $\dot{M}_\text{Edd}$ is the Eddington accretion rate\footnote{$\dot{M}_\text{Edd}=(4\pi GM_{\rm p} m_{\rm{p}})/(\sigma_{\rm{T}}\eta c)$, where $\sigma_{\rm{T}}$ is the Thomson cross section and $m_{\rm{p}}$ is the proton mass.} assuming a radiative efficiency $\eta=0.1$. While most of the observational evidence suggests $\alpha\sim0.1$ (see \citealt{King2007} and references therein),  magneto-hydrodynamical (MHD) simulations suggest $\alpha\sim0.01$ \citep[see, e.g.][]{Hawley2011}, which we also choose. Furthermore, we will show in Section \ref{Sec4} and Appendix \ref{AppB} that the most direct consequence of a different $\alpha$ is whether a gap/cavity opens for a particular set of system parameters. For higher $\alpha$, we expect faster viscous refilling leading to less likelihood of gap/cavity and more likelihood of stronger Type-I torque. Therefore, the choice of a lower $\alpha$ makes our results in Section \ref{Sec5} conservative estimates. 

\begin{figure}
\centering{\includegraphics[width=1\linewidth]{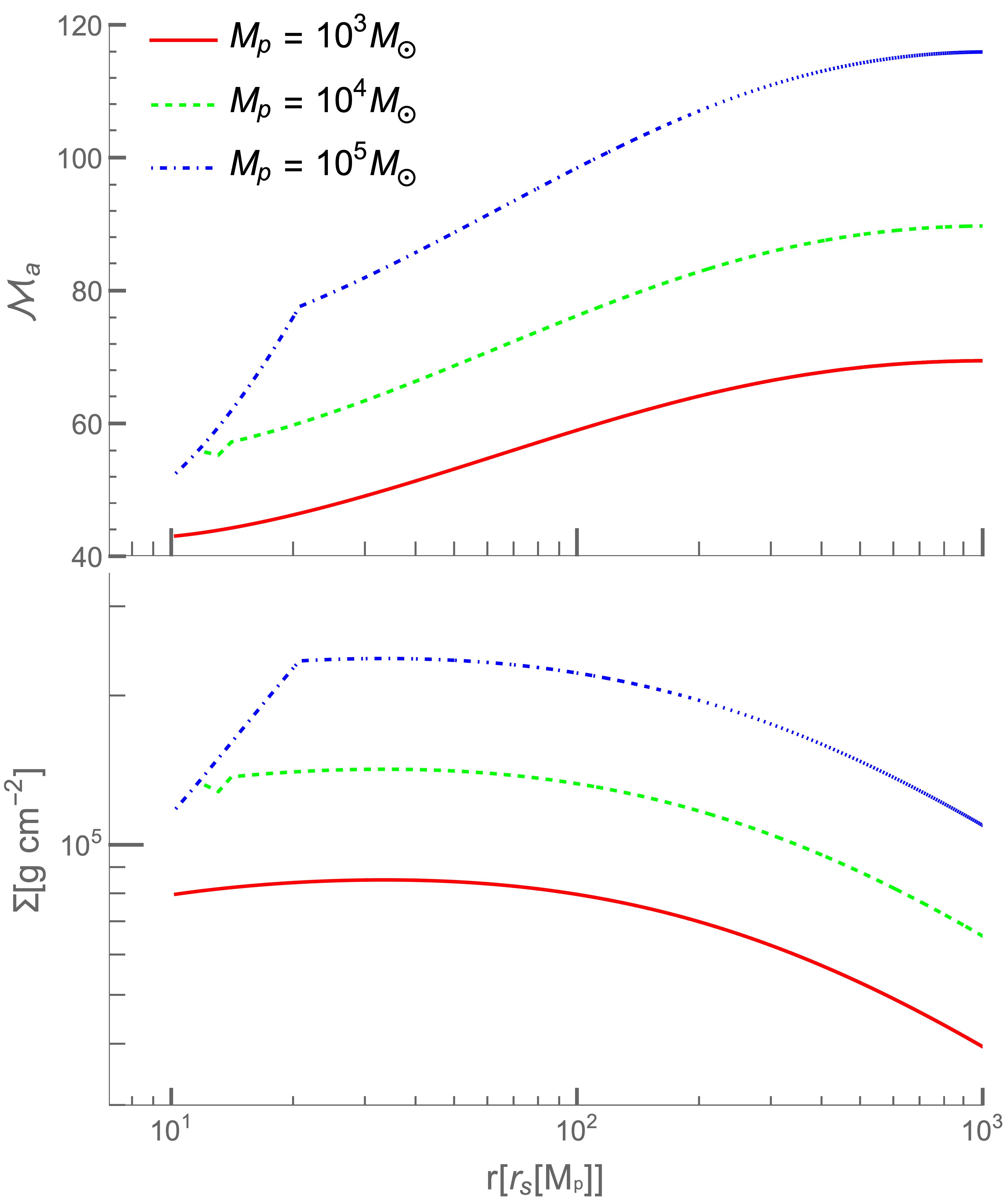}}
\caption{Fiducial radial profiles of the Mach number $\mach$ and surface density $\Sigma$ for three different $\alpha$ discs with central BH masses $10^3$ (solid red), $10^4$ (dashed green), and $10^5~\MSun$ (dot-dashed blue).}
\label{Fig3} 
\end{figure}

In Fig.~\ref{Fig3}, we show the $\Sigma$ and $\mach$ radial profiles with respect to the Schwarzschild radius of the central BH. Note that the profiles have a relatively weak dependence on $r$ and $M_{\rm{p}}$, and they can be defined by fiducial values:

\begin{align}\label{Eq9}
    \machz&=80,\nonumber\\
    \Sigma_0&=2\times10^5~\text{g cm}^{-2}.
\end{align}

The Mach number depends on the accretion rate, and typically we expect higher Mach numbers (or thinner, denser discs) for more massive central BHs as per Fig.~\ref{Fig3}. The volumetric density $\rho=\Sigma/(2h)\approx10^{-5}$ g cm$^{-3}$ at $r_\text{ISCO}$ is around one order-of-magnitude higher than the expected value from the survey of Seyfert galaxies (as inferred from fig.~8 in \citealt{JiangJ2019}). Therefore, our disc solutions are reasonable. Moreover, we will show in the following section that the gas torques directly scale with the accretion rate $\dot{M}$; hence, exact values of $\mach$ and $\Sigma$ are not crucial. In the next section, we study the evolution of a BH binary due to being embedded in a gas accretion disc.

\section{Migration}\label{Sec4}

We expect either one or both components of a BHB embedded in a gas disc to experience torques via an exchange of angular momentum with nonaxisymmetric perturbations \citep{Goldreich1980, Ward1997}. Here we introduce the various regimes of gas response to an embedded perturber (and resulting torques), which depend primarily on the binary mass ratio. In all cases we define torques in the limit that the binary mass is much greater than the enclosed disc mass. 

In the limit of small secondary mass ($q\ll 1$), the Type-I torque exerted by the gas on the secondary can be analytically expressed by

\begin{equation}\label{Eq10}
    \Gamma_\text{lin}=\Sigma r^4\Omega^2q^2\mach^2,
\end{equation}

\noindent assuming the gas is isothermal (and neglecting factors of a few which arise due to disc gradients), which has been tested in 3D simulations by \citet{Tanaka2002}. 

For higher mass ratios, the secondary will begin to repel the gas along its orbit as it sheds its orbital angular momentum by launching spiral density waves,  carving out a low-density region, referred to as the `gap'. It is assumed that, in the limit that the secondary mass is small compared to the disc mass, the secondary is not able to perturb directly the gas disc anymore, so that its orbit is tied to the viscous diffusion in the disc \citep{Lin1986}. Therefore, in the latter regime, which is termed `Type II', the torque on the secondary is equal to the viscous torque

\begin{equation}\label{Eq11}
    \Gamma_{\rm vis} = \dot{M} r^2 \Omega=3\pi\alpha\Sigma r^4\Omega^2\mach^{-2}.  
\end{equation}

When the secondary BH is more massive than the local disc, torques weaken by a factor that depends on disc mass \citep{Syer1995}. This becomes a substantially weaker torque than the Type-I regime for higher mass ratios. However, recent works show that the Type-II torque can be more accurately expressed by the linear torque with a correction factor that takes into account the gap structure \citep{Fung2014,Kanagawa2018}:

\begin{equation}\label{Eq12}
    \Gamma_{\text{gap}}=\frac{1}{1+0.04(q^2\mach^5\alpha^{-1})}\Gamma_{\text{lin}}.
\end{equation}

Under standard assumptions, Type-II torques are weaker than the Type-I torque. However, numerical simulations show that once a gap is carved gas can still flow through it from the disc, generating new contributions to the net torque on the secondary. These  non-linear gas dynamics effects can even lead to changes in the sign of the net torque \citep{Duffell2014}. Since their description is beyond the capabilities of an analytical models such as that employed here, we will neglect them. The mass ratio above which the Type-II regime occurs depends non-trivially on the disc parameters. We address this nuance in Appendix~\ref{AppB}.

For mass ratios above $q=0.04$ \citep{DOrazio2015}, torques exerted by the binary typically clear out the inner disc gas, carving out a central cavity.\footnote{Thermodynamics and initial conditions of the binary may affect the mass ratio for gap-to-cavity transitions, including at which separation it forms (as shown by \citealt{Lima2020} for a fixed mass ratio of $q=0.05$ and various disc parameters).}  Gas still flows near the BHs (via streams, \citealt{Farris2014}) and can torque the binary with a strength that usually depends on the accretion rate into the cavity \citep{Munoz2019,Moody2019,Duffell2020,Dittmann2022}. Circumbinary disc (CBD) simulations that directly measure the steady-state gas torque on the binary (not only the secondary) typically express it in terms of the viscous torque $\Gamma_\text{CBD}=\xi \Gamma_\text{vis}$, where we introduce a fudge factor $\xi$, which depends on the binary mass ratio or disc parameters (see, e.g. \citealt{Duffell2020}).

\begin{figure}
\centering{\subfloat[]{\includegraphics[width=1\linewidth]{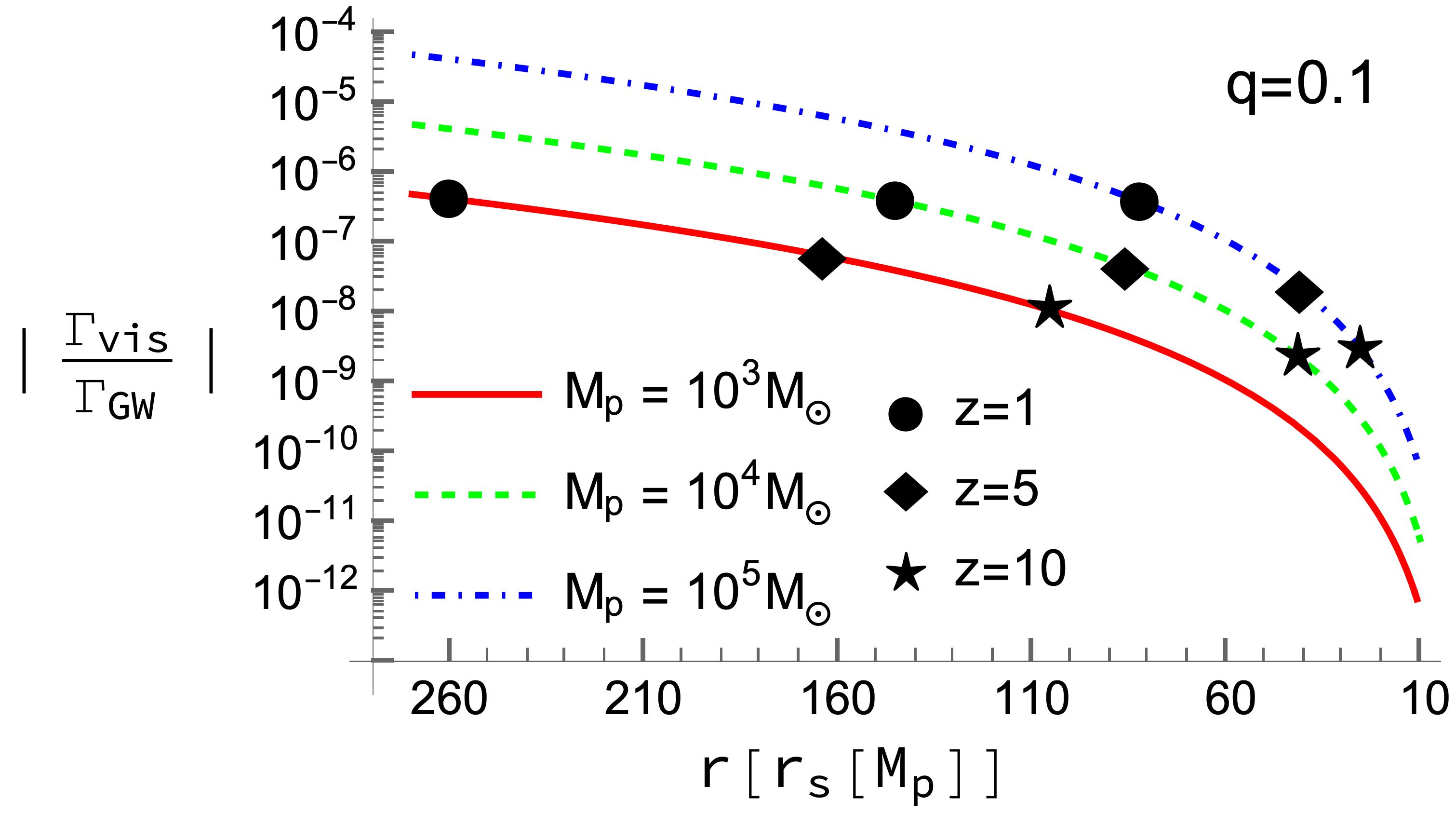}}\\
\subfloat[]{\includegraphics[width=1\linewidth]{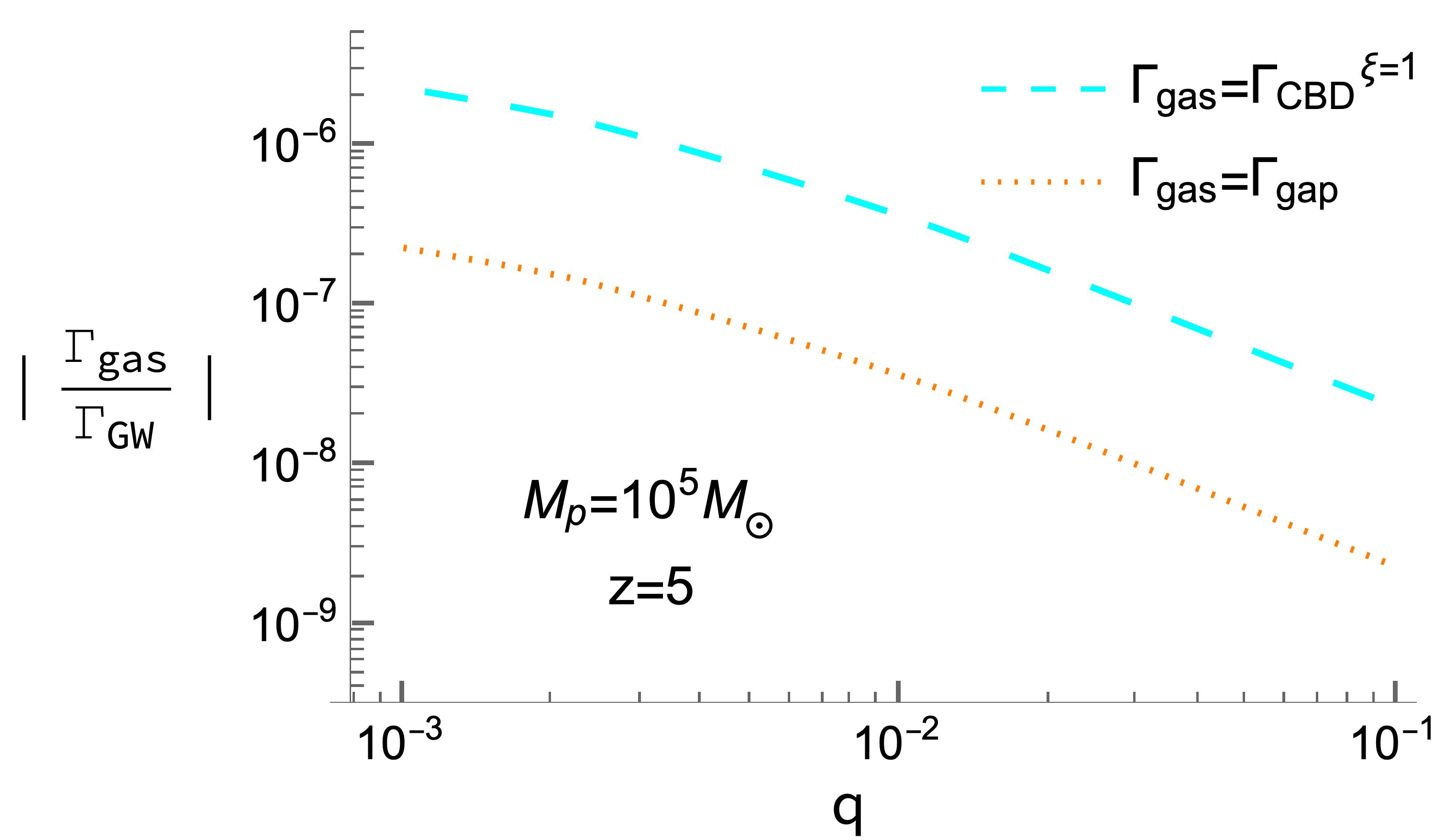}}}
\caption{(a) $\Gamma_\text{vis}$ normalized by $\Gamma_\text{\scriptsize{GW}}$ as a function of binary separation for three different primary masses $M_{\rm{p}}=10^3\text{ (solid red)},~10^4\text{ (dashed green)},$ and $10^5~\MSun\text{ (dot-dashed blue)}$ with a mass ratio $q=0.1$. We also mark at which separation each system will enter the LISA band for three redshifts $z=1$ ($\bullet$),~$z=5$ ({\tiny $\blacklozenge$}), and $z=10$ ({\scriptsize $\bigstar$}). (b) Type-II/cavity gas torques $\Gamma_{\text{gap}}$ (dotted orange) and $\Gamma_{\text{CBD}}$ (for $\xi=1$, long-dashed cyan) with respect to $\Gamma_\text{\scriptsize{GW}}$ at the LISA entry as a function of the mass ratio for a primary mass of $10^5~\MSun$ and redshift $z=5$. In both panels, we employ the full Mach number and surface density radial profiles shown in Fig.~\ref{Fig3}. In the top panel, $\Gamma_\text{vis}/\Gamma_\text{\scriptsize{GW}}\propto\dot{M}\propto M_{\rm p}$ for the fixed mass ratio and separation normalized by the Schwarzschild radius of the primary BH.}
\label{Fig4} 
\end{figure}

In both the Type-II and cavity regimes, the torque on the secondary or the binary scales with the viscous torque (this is less obvious with the gap-corrected expression in Eq.~\ref{Eq12}, though we show this in Fig.~\ref{Fig4}). In other words, torques scale with the disc's accretion rate times some factor $\xi$ that is sensitive to disc characteristics. Simulations in both regimes find that the torque strength and direction (i.e. the value and sign of $\xi$) has a nontrivial dependence on disc parameters, particularly the Mach number and accretion rate/viscosity \citep{Duffell2014,Duffell2020,Tiede2020,Dittmann2022} as well as numerical sink prescriptions \citep{Dittmann2021}. For example, the parameter study by \citet{Duffell2020} (which covers the gap-to-cavity transition with high-resolution, 2D hydrodynamic simulations) finds that for $\mach=10$ and mass ratios above $q=0.05$,  $\xi\approx0.6$. For smaller $q$, or for higher values of viscosity or Mach number explored in similar studies, $|\xi|\lesssim2$ (as inferred from Fig.~3 of \citealt{Dittmann2022}). For higher Mach numbers (or equivalently smaller aspect ratios), torques become increasingly negative \citep{Tiede2020}. Together, these studies find that $0.1 \lesssim |\xi| \lesssim 2$ and suggest that even higher values of $\xi$ may arise in more realistic discs where $\mach\sim100$ (see Fig.~\ref{Fig3}), provided that the scaling of torque with Mach number persists beyond the currently explored regime. Therefore, we will consider two limiting cases to bracket the possibilities for both Type-II and cavity regime torques:

\begin{equation}\label{Eq14}
    \Gamma_\text{gas}=\xi \Gamma_\text{vis},
\end{equation}

\noindent with $\xi=0.1$ and $\xi=10$. Both Type-II and cavity torques are affected by the gas diffusion into the gap and cavity, respectively. The uncertainty in $\xi$ arises from a similar physical process, namely the nonlinear gas flow around the BHs. For this reason, we use the same fudge factor in $\Gamma_{\rm CBD}$ and Eq.~\eqref{Eq14}.

For our system parameters, we found that we always expect a gap/cavity to open and influence the binary evolution (see Appendix~\ref{AppB} for a detailed discussion) and, therefore, we are in the Type-II/cavity regime with the effective gas torque $\Gamma_\text{gas}$. Furthermore, most of the mass ratios of our interest are less than $0.04$, where the gap-to-cavity transition happens. Hence, we are primarily in the Type-II regime rather than the cavity regime.

The binaries are primarily inspiraling due to the emission of GWs, which for comparison can be expressed as an effective GW torque on the secondary BH,

\begin{equation}\label{Eq15}
    \Gamma^\text{sec}_\text{\scriptsize{GW}}=\frac{1}{2}qM_{\rm{p}}r\dot{r}_\text{\scriptsize{GW}}\Omega,
\end{equation}

\noindent or on the binary:\footnote{By replacing the secondary mass in Eq.~\eqref{Eq15} by the reduced mass $\mu\equiv qM_{\rm{p}}/({1+q})$ of the binary.}

\begin{equation}\label{Eq16}
    \Gamma^{\text{bin}}_\text{\scriptsize{GW}}=\frac{1}{2}\frac{qM_{\rm{p}}}{1+q}r\dot{r}_\text{\scriptsize{GW}}\Omega.
\end{equation}

Since we are considering mass ratios in the range $10^{-3}$--$10^{-1}$, $1+q \simeq 1$ and $\Gamma^{\text{bin}}_\text{\scriptsize{GW}}\approx \Gamma^{\text{sec}}_\text{\scriptsize{GW}}$ (even for $q=0.1$). Thus, we will simply use $\Gamma_\text{\scriptsize{GW}}$ to denote the effective GW torque and use the expression given in Eq.~\eqref{Eq15}. In the top panel of Fig.~\ref{Fig4}, we show $\Gamma_{\text{vis}}$ with respect to $\Gamma_{\text{GW}}$ for three primary BH masses, $10^3$, $10^4$, and $10^5~\MSun$, with a binary mass ratio $q=0.1$. We also indicate at which rest-frame separation respective binaries enter the LISA band at different redshifts. In the bottom panel, we show Type-II/cavity gas torques, $\Gamma_{\text{gap}}$ and  $\Gamma_{\text{CBD}}$ (for $\xi=1$, i.e. $\Gamma_{\text{vis}}$), with respect to $\Gamma_\text{\scriptsize{GW}}$ at the LISA entry as a function of the mass ratio. We use the full Mach number and surface density radial profiles shown in Fig.~\ref{Fig3}.

Fig.~\ref{Fig4} suggests that, irrespective of the mass ratio, $\Gamma_\text{gap}\approx0.1\Gamma_\text{vis}$.\footnote{A similar torque strength is measured in simulations for low $q$ by \citet{Derdzinski2019}, who also consider gas effects on GW sources.} Also, when IMBHBs with $q=0.1$ at $z=5$ enter the LISA band, we have $\Gamma_\text{vis}/\Gamma_\text{\scriptsize{GW}}\approx10^{-7}$, which further implies $\Gamma_\text{gap}/\Gamma_\text{\scriptsize{GW}}\approx10^{-8}$. The ratio between the gas torque and the GW torque reduces exponentially as binaries approach merger. However, this does not directly tell us if we can detect gas-related observables using LISA, because in principle even small deviations are detectable in events with high SNR (as we will show in Section~\ref{Sec5}).

In the following section, we discuss the gas imprints on the GW waveform and recovered binary parameters, that result from a thin gas accretion disc.

\section{Observables and critical parameters}\label{Sec5}

We first compute the GW phase shift due to the presence of an $\alpha$-disc and then extrapolate it to super-Eddington rates. Then we discuss the statistical significance of gas-induced bias on the estimate of the binary chirp mass. We conclude this section by computing the limiting values of the surface density and the speed of sound required for an accretion disc in our model to make dephasing due to gas detectable by LISA.

\subsection{Dephasing due to gas}\label{Sec5.1}

The phase shift $\delta\phi$ is the absolute difference of the GW phase with or without the influence of a gas disc. The total phase of GW in radians for a circular binary evolving between two separations $r_{\rm min}$ and $r_{\rm max}$ is given by

\begin{equation}\label{Eq17}
    \Phi=2\pi\int^{r_{\rm min}}_{r_{\rm max}} {\rm d}r\frac{{f_{\rm{r}}}}{\dot{r}},
\end{equation}

\noindent where here $\dot{r}$ is a general orbital evolution rate, which is $\dot{r}_\text{\scriptsize{GW}}$ in a vacuum. It can also be interpreted as the total number of orbits spent by a BHB between separations $r_{\rm max}$ and $r_{\rm min}$ with the relation $N_\text{orb} = |\Phi|/4\pi$.

The presence of gas affects the orbital evolution, and up to leading order we can take $\dot{r}=\dot{r}_\text{\scriptsize{GW}}+\dot{r}_\text{gas}$ with\footnote{Derived by equating the gas torque to the time derivative of the angular momentum of the reduced mass due to gas and ignoring mass accretion term.} $\dot{r}_{\text{gas}}=2 \Gamma_\text{gas}(1+q)(qM_{\rm{p}})^{-1}r^{1/2}(GM_\text{tot})^{-1/2}$, given that  $\dot{r}_\text{gas}\ll\dot{r}_\text{\scriptsize{GW}}$ as per Fig.~\ref{Fig4}. Therefore, the phase shift due to gas in the LISA observation window becomes

\begin{equation}\label{Eq18}
    \delta\phi=|\Phi(\dot{r}_\text{\scriptsize{GW}})-\Phi(\dot{r}_\text{\scriptsize{GW}}+\dot{r}_\text{gas})|\approx2\pi\int^{r_{\rm f}}_{r_{\rm i}} {\rm d}r{f_{\rm{r}}}\frac{\dot{r}_\text{gas}}{\dot{r}_\text{\scriptsize{GW}}^2},
\end{equation}

\noindent where we have used $\dot{r}_\text{gas}\ll\dot{r}_\text{\scriptsize{GW}}$ to simplify this expression.

For LISA to observe this phase shift, we compute a measure of the detectability of the dephasing (which is analogous to the SNR of the difference between the vacuum and gas-affected waveforms), following \citet{Kocsis2011}:

\begin{equation}\label{Eq19}
    \text{SNR}_{\delta\phi}=\sqrt{2 \cdot 4\int^{{f}_{\text{f}}}_{{f}_{\text{i}}}{\rm d}{f}^\prime\frac{2h_{\rm{c}}^2({f}^\prime)(1-\cos{\delta\phi})}{S_{\rm{t}}({f}^\prime){f}^{\prime2}}}.
\end{equation}

As before, we use $\text{SNR}_{\delta\phi}\geq8$ as a phase shift detectability threshold. Four parameters -- $M_{\rm{p}},~q,~z,$ and $\dot{M}$ -- determine how $\Gamma_\text{gas}$ varies throughout the binary evolution. To better understand the dependence on each parameter, we fix one of these four variables at a time, to quantify the dephasing due to gas in the LISA band for its four-year observation window. While the direction of the gas torque would decide whether a binary gains or loses orbits, the absolute change in the number of orbits will be the same as per Eq.~\eqref{Eq18}. One can also convert the phase shift to the percentage change in the total number of orbits by

\begin{equation}\label{Eq20}
    \delta N[\%]=100\times\frac{\delta N_{\text{LISA}}}{N_\text{LISA}},
\end{equation}

\noindent where $\delta N_{\text{LISA}}=\delta\phi/4\pi$ is the change in the total number of orbits and $N_\text{LISA}$ is given in Fig.~\ref{Fig2}.

\subsubsection{Varying MBH primary masses, binary mass ratios, and redshifts}\label{4.1.1}

\begin{figure}
\centering
\subfloat[]{\includegraphics[width=1\linewidth]{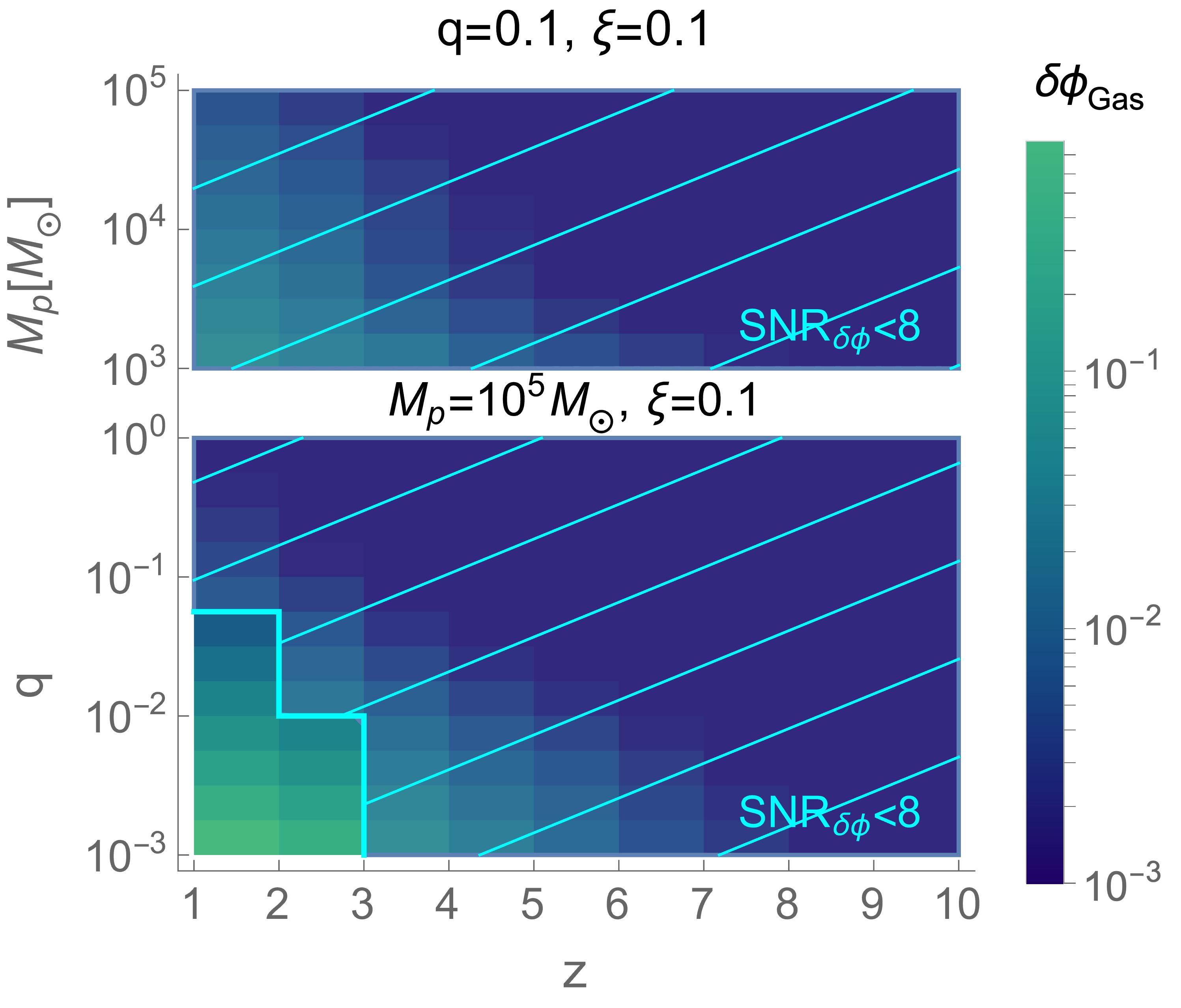}}\\
\subfloat[]{\includegraphics[width=1\linewidth]{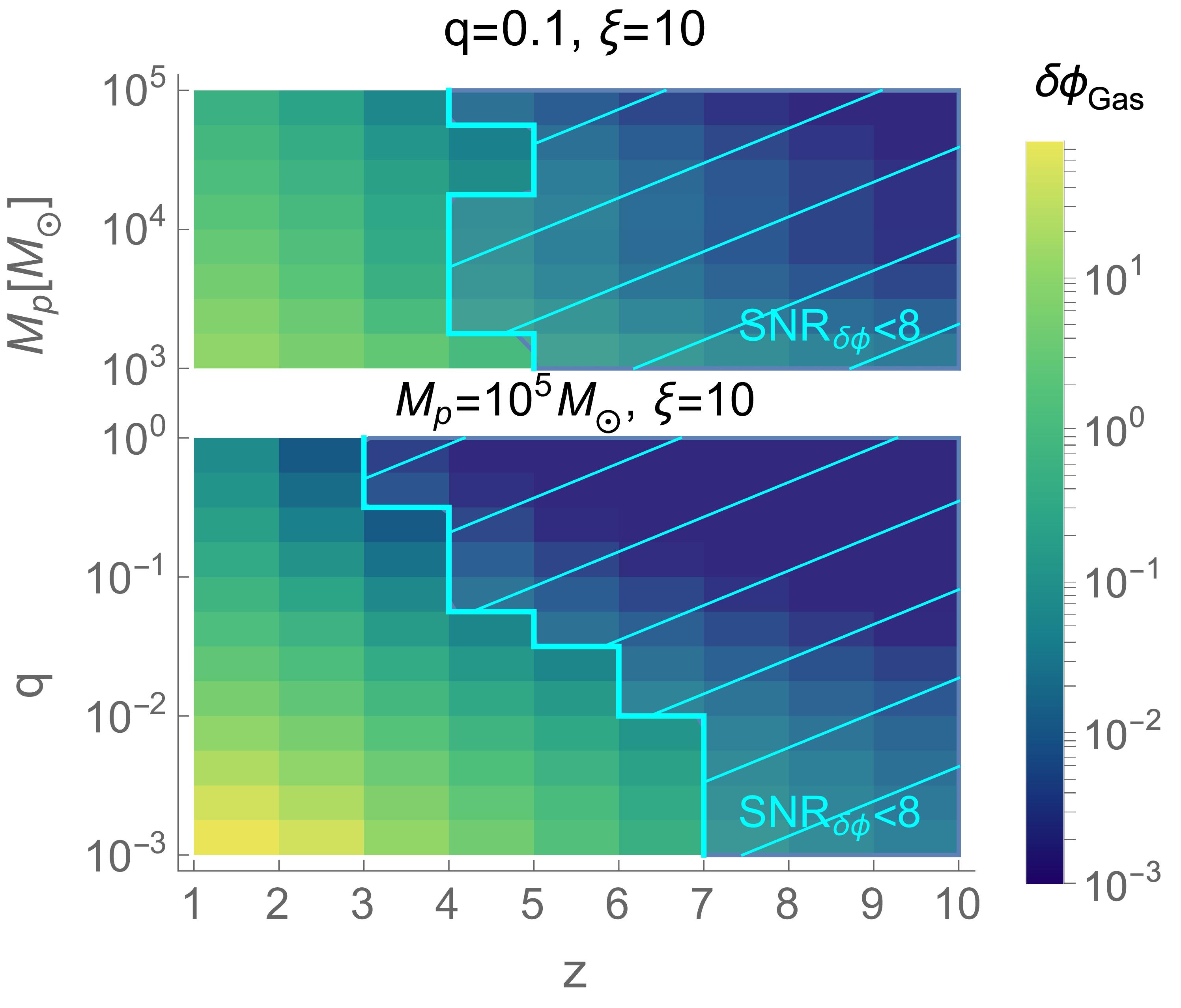}}\\
\caption{For the same binary parameters as in Fig.~\ref{Fig2}, accumulated phase shifts $\delta\phi_\text{gas}$ during the LISA observation window, for the fudge factors (a) $\xi=0.1$ and (b) $\xi=10$, $\mach=80$ and $\Sigma=2\times10^5 \text{ g cm}^{-2}$. Both panels have a LISA non-detectability region (SNR$_{\delta\phi}<8$) shaded with diagonal cyan lines. For reference, we have extended mass ratios up to unity in both pair of plots in the bottom panels.}
\label{Fig5} 
\end{figure}

We compute phase shifts for the BH binary parameters shown in Fig.~\ref{Fig2}, assuming the fiducial values for the Mach number and surface density given in Eq.~$\eqref{Eq9}$. In Fig.~\ref{Fig5}, we show the dephasing in the LISA band, which we denote with $\delta\phi_\text{Gas}$, for both $\xi=0.1$ and $\xi=10$. For $\xi=0.1$, we can infer that our modelled gas effects are not strong enough to be detected in signals from sources with mass ratios $\gtrsim0.05$ and redshift $\gtrsim3$. For $\xi=10$, binaries with $q\lesssim0.1$ need to be at redshift $z\lesssim5$ to have SNR$_{\delta\phi}\geq8$ and for low mass ratios, LISA can observe phase shifts up to $z\sim7$. The range of detectable dephasing for $\xi=0.1$ is $\delta\phi\sim0.014-0.96$ radians\footnote{Note that the each tile is an average value between dephasing corresponding to two adjacent set of parameters, hence the maximum value seen on the plot is less than 0.96 radians.}, which corresponds to maximum $\delta N[\%]\sim10^{-4}$. For $\xi=10$, observable phase shifts are $\delta\phi\sim0.029-96$ radians with maximum $\delta N[\%]\sim10^{-2}$. In other words, the signal from the binary will be shifted by a fraction of an orbit to several orbits, depending on the gas torques. For high SNR sources, such phase shifts are detectable assuming we have sufficiently accurate waveforms (we discuss the importance of this assumption in Section~\ref{Sec6}).

\subsubsection{Varying the MBH primary masses, the binary mass ratios, and the accretion rates}\label{4.1.2}

We calculate dephasing for similar binaries as in the previous section, but this time, we only consider redshift $z=10$. We are interested in studying binaries at high redshift because the remnant BH could be a building block of supermassive quasars observed at $z\sim7$, as discussed in the introduction. However, Fig.~\ref{Fig5} shows that, for our Eddington-limited $\alpha$-disc, dephasing is too weak to be detectable for redshifts higher than $z\sim7$. Assuming that the gas torque scaling with $\dot{M}$ continues into the super-Eddington regime, we strengthen the phase shift by increasing $\dot{M}$ from $0.1\dot{M}_\text{Edd}$ to $100\dot{M}_\text{Edd}$. This is equivalent to increasing the surface density or decreasing the Mach number (see Appendix \ref{AppA} for the $\dot{M}$ dependence on disc parameters). High accretion rates are indeed expected at high redshift, as BH seeds may require either continuous Eddington accretion or episodes of super-Eddington accretion to reach the mass of observed quasars at $z \sim 7$ \citep[see, e.g.][]{Mayer2013,Mayer2019}. Furthermore, local observations of active dwarf galaxies are consistent with super-Eddington rates \citep[see, e.g.][]{Mezcua2017}, which should be more likely at high redshifts.

\begin{figure}
\centering
\subfloat[]{\includegraphics[width=1\linewidth]{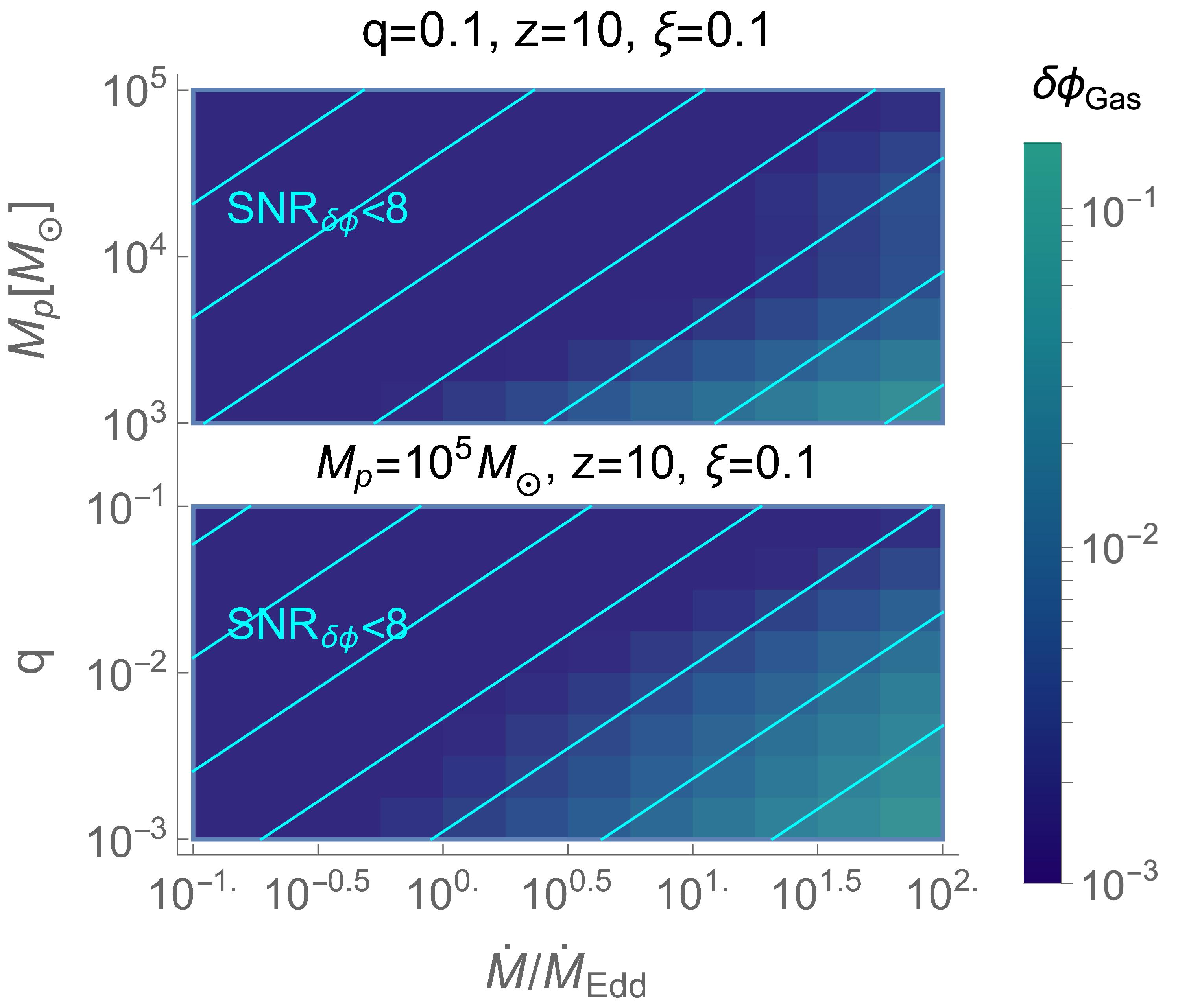}}\\
\subfloat[]{\includegraphics[width=1\linewidth]{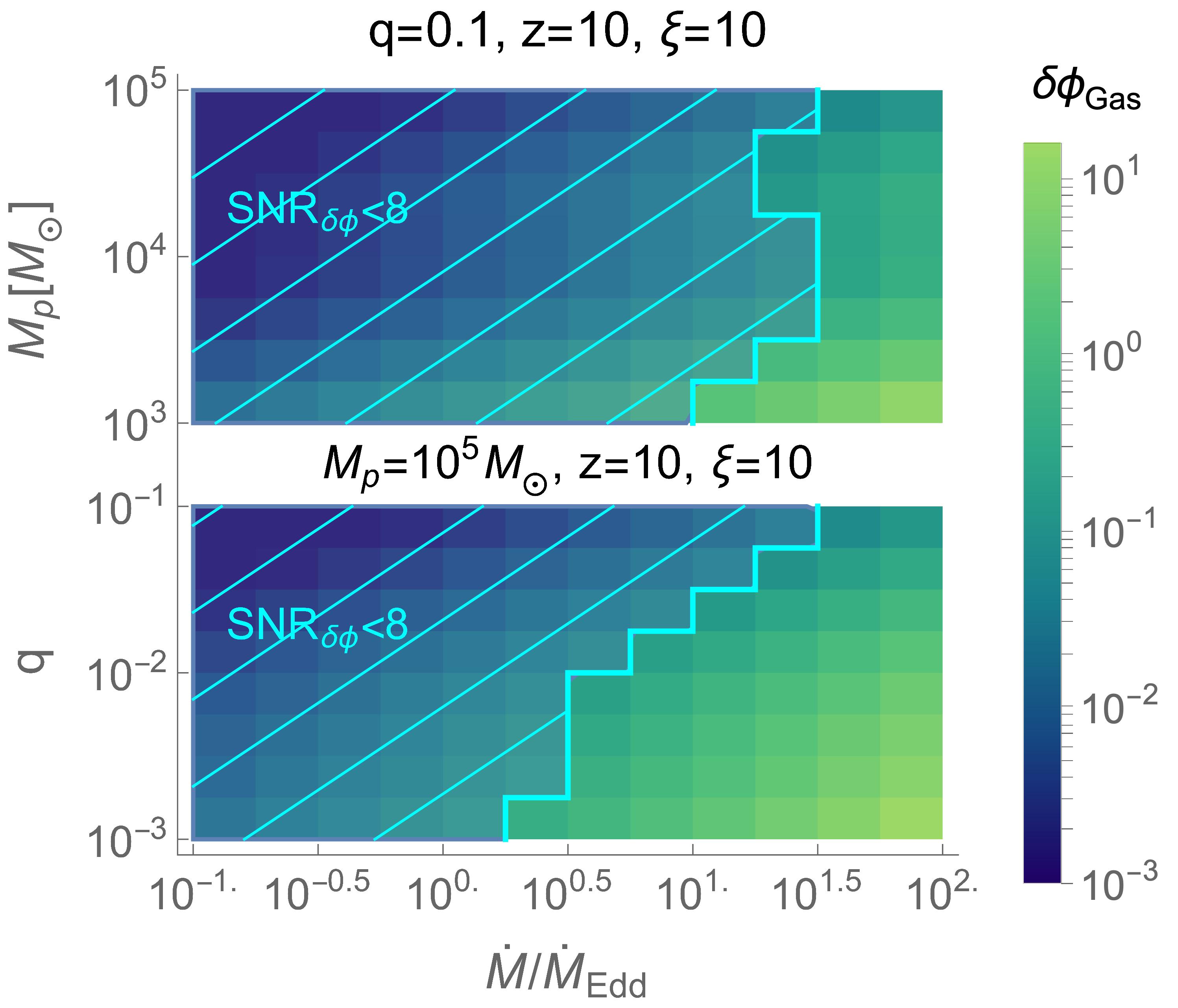}}
\caption{Same as Fig.~\ref{Fig5}, but at a fixed redshift ($z = 10$) and varying the accretion rate.}
\label{Fig6} 
\end{figure}

We show our results in the form of colour-coded density plots in Fig.~\ref{Fig6} for both $\xi=0.1$ and $\xi=10$ fudge factors. Phase shifts are only observable for $\xi=10$, for which they fall in the range of $\delta\phi\sim0.051-15$ radians with maximum $\delta N[\%]\sim10^{-2}$. We can infer from this figure that LISA can only observe dephasing for high-redshift binaries only if we have super-Eddington accretion rates. We caution that the disc geometry changes in the super-Eddington regime (see, e.g. the slim disc solution by \citealt{Abramowicz1988} or \citealt{Jiang2019}, wherein they reach $\dot{M}=150\dot{M}_\text{Edd}$) and the resulting torques may change as well. For instance, in the quasi-Keplerian isothermal slim disc model, the aspect ratio $h/r$ is larger by a factor of $\sqrt{6}$ than the thin disc.\footnote{It follows from the relation $\Omega^2h^2=6(p/\rho)=6c^2_{\rm s}$ in \citet{Abramowicz1988}, where $p$ is the total pressure.} The slim disc viscous torque $\Gamma_{\rm vis,slim}=4\pi r^2\alpha ph$ is then smaller by a factor of $2/3\sqrt{6}$ than the thin disc $\Gamma_{\rm vis}$ for the same viscosity prescription, $\alpha$, and $\dot{M}$, which suggests our estimates in the super-Eddington regime could be off by a factor of order unity. Moreover, an additional cooling mechanism may occur in the slim disc model that would reduce $h/r$, and the corresponding torque needs to be determined by a high-resolution simulation. Nevertheless, these estimates could give us an order-of-magnitude idea about the gas-induced phase shift. 

This section shows that the dephasing is detectable for binaries embedded in a gas disc, when in the sub-Eddington regime. If $\xi=0.1$, then phase shifts are detectable up to $z\sim3$ but only for small mass ratios $q\sim10^{-3}$. For $\xi=10$, dephasing is observable until $z\sim5$ for $q\sim0.1$ and $z\sim7$ for $q\sim10^{-3}$. In other cases, it is not detectable because either the inspiral is too fast or the gas torques are too weak. For our fiducial model, dephasing is undetectable for high-redshift (e.g. at $z=10$) IMBHBs. If these binaries are accreting at super-Eddington rates and the scaling of gas torques with $\dot{M}$ still applies in this regime, then dephasing can be detectable.

\subsection{Bias in the chirp mass estimate}\label{Sec5.2}

This section explores the bias in the recovered chirp mass $\mathcal{M}$ from the signal data if we assume a vacuum circular binary model and ignore any gas effects. We are interested in performing this analysis to gauge how careful one needs to be while running parameter estimation pipelines to extract binary parameters without taking into account the presence of gas. Similar work has been presented by \citet{Chen2019} and \citet{Antoni2019} for LIGO sources and by \citet{Caputo2020} for both LISA-LIGO multi-band sources and LISA-only IMBHBs. While \citet{Caputo2020} consider a similar mass range as in this work, they only consider redshifts $z=0.1$ and $z=0.5$. Furthermore, they study dephasing from gas accretion on to BHBs, not due to torques. Therefore, our analysis provides a novel estimate of bias in the chirp mass for IMBHBs under the influence of gas torques at high redshifts. 

Under the stationary phase approximation, we expect gas to only affect the phase (with phase shift $\delta\phi$) and not the amplitude strain $h$. However, gas can perturb the frequency evolution rate, which can be interpreted as a change in the chirp mass. The rest-frame time derivative of the rest-frame frequency $f_{\rm r}$ is

\begin{equation}\label{Eq21}
    \dot{f}_{\rm{r}}=\frac{1}{\pi}\left(\frac{GM_\text{tot}}{r^5}\right)^\frac12\left(-\frac32\right)\dot{r}.
\end{equation}

Plugging Eq.~\eqref{Eq21} into Eq.~\eqref{Eq4} yields

\begin{equation}\label{Eq22} 
    \mathcal{M}=\frac{c^3}{G}\left(-\frac{5}{64}(GM_\text{tot})^{-\frac43}r^3\dot{r}\right)^{\frac35}.
\end{equation}

The interesting part here is that the radial dependence of the chirp mass is $\mathcal{M}\propto r^3\dot{r}$. For the case of a circular BH binary evolution in a vacuum, where $\dot{r}=\dot{r}_\text{\scriptsize{GW}}\propto r^{-3}$, the apparent chirp mass is separation-independent. In the presence of gas, however, we have $\dot{r}=\dot{r}_\text{\scriptsize{GW}}+\dot{r}_\text{gas}$, where, under the assumptions of this work, $\dot{r}_\text{gas}\propto r$ due to $\Gamma_{\rm gas}$. Therefore, we expect the apparent chirp mass to be separation-dependent (and, consequently, time-dependent). In the presence of gas, the apparent chirp mass at a given separation $r$ is

\begin{equation}\label{Eq23}
    \mathcal{M}_\text{gas}(r)=\mathcal{M}_\text{tr}\left(1+\frac{\dot{r}_\text{gas}}{\dot{r}_\text{\scriptsize{GW}}}\right)^{\frac{3}{5}},
\end{equation}

\noindent where $\mathcal{M}_\text{tr}$ is the true chirp mass (i.e. for a binary in vacuum).

By affecting the number of orbits a binary spends at each frequency, the gaseous disc affects the characteristic strain $h_\text{c,gas}\equiv h_\text{c}e^{i\delta\phi}$. Assuming $h_\text{c,gas}$ as an observed characteristic strain and the vacuum model as the underlying model, we are interested in calculating the bias in chirp mass estimate, which illustrates the importance of the systematic error due to not considering gas with respect to the statistical uncertainty in the recovery of the source chirp mass. Because of $h^2_\text{c,gas}=h^2_\text{c}$, we have the same SNR of the observed signal as in Eq.~\eqref{Eq8}.\footnote{This is valid in the limit of $\Gamma_\text{gas}/\Gamma_{\text{\scriptsize{GW}}}\ll1$.}

To demonstrate this effect, we pick a high SNR source for which dephasing is detectable in Fig.~\ref{Fig5}. Our fiducial binary parameters are $M_{\rm{p}}=10^5~\MSun$, $q=0.1$, and $z=1$. The fiducial gas parameters are $\mach=80$ and $\Sigma=2\times10^5~\text{g cm}^{-2}$, and we consider the stronger gas torque, $\Gamma_\text{gas}=10\Gamma_\text{vis}$.

The frequency evolution rate $\dot{f}_{\rm{r}}$ depends upon $\dot{r}$ according to Eq.~\eqref{Eq21}. Hence, we can express the gas-effective $\dot{f}_{\rm{r,gas}}$ in terms of the vacuum $\dot{f}_{\rm{r,GW}}$ (Eq.~\eqref{Eq21} for $\dot{r}=\dot{r}_\text{\scriptsize{GW}}$) as

\begin{equation}\label{Eq24}
    \dot{f}_{\rm{r,gas}}=\dot{f}_{\rm{r,GW}}\left(1+\frac{\dot{r}_\text{gas}}{\dot{r}_\text{\scriptsize{GW}}}\right).
\end{equation}

We can quantify the percentage difference between $\dot{f}_{\rm{r,gas}}$ and $\dot{f}_{\rm{r,GW}}$ by defining

\begin{equation}\label{Eq25}
    \delta \dot{f}_{\text{r,Gas}}[\%]=100\times\frac{\dot{f}_{\rm{r,gas}}-\dot{f}_{\rm{r,GW}}}{\dot{f}_{\rm{r,GW}}}=100\frac{\dot{r}_\text{gas}}{\dot{r}_\text{\scriptsize{GW}}}.
\end{equation}

\begin{figure}
\centering
{\includegraphics[width=1\linewidth]{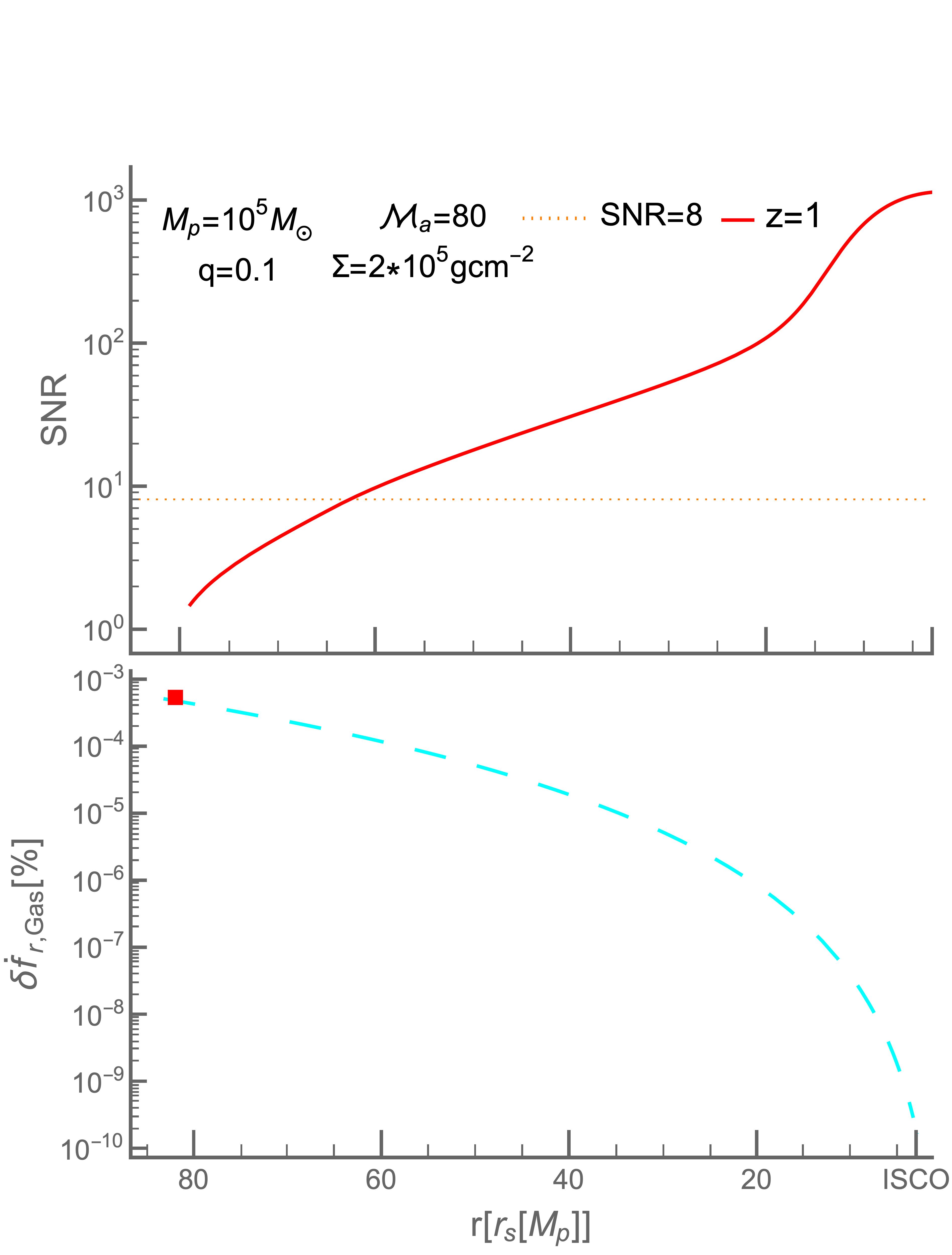}}
\caption{The effect of gas on the frequency evolution rate $\dot{f}_{\rm{r}}$ of a GW waveform due to gas, quantified by $\delta \dot{f}_{\text{r,Gas}}[\%]$ (bottom panel) for a binary with $M_{
\rm p}=10^5~\MSun$, $q=0.1$, and $z=1$, with gas parameters $\mach=80$ and $\Sigma=2\times10^5$~g~cm$^{-2}$. We also mark with a red square when this binary enters the LISA band. Furthermore, the top panel shows the accumulated SNR of the event throughout the binary evolution until $r_{\rm ISCO}$.} 
\label{Fig7} 
\end{figure}

In Fig.~\ref{Fig7}, we show $\delta \dot{f}_{\text{r,Gas}}[\%]$ in terms of the binary radial separation for our fiducial system parameters. This figure illustrates that gas effects on the waveform become weaker while the accumulated SNR becomes higher towards the merger. Therefore, if we consider the whole signal data, the louder late-stage inspiral part, wherein
we expect near-vacuum quantities, will further suppress any non-vacuum signatures.

To compute the bias in chirp mass estimate, we use the Fisher information matrix analysis \citep{Cutler2007}. To simplify the analysis, we assume that the strain is solely a function of the chirp mass,\footnote{This should be reasonable as the chirp mass is the leading order effect in the BHB evolution \citep[see, e.g.][]{Sathyaprakash2009}.} i.e. $h\equiv h(\mathcal{M})$, and the analysis is performed after collecting full data for the given binary in the LISA observation window of four years. The late-stage inspiral part of the data would allow us to get the true chirp mass ($\mathcal{M}_\text{tr}$), as non-vacuum effects will be heavily suppressed. This also reduces the Fisher matrix to one element. Taking the Gaussian distribution for the parameter $\mathcal{M}$, an inherent assumption of the Fisher analysis, the mean of the distribution would be the true chirp mass:

\begin{equation}\label{Eq26}
    \mu_{\mathcal{M}}=\mathcal{M}_\text{tr}={(M_{\rm{p}}^2 q)^\frac35}/{M_\text{tot}^\frac15}.
\end{equation}

The inverse of the variance is the only element of our Fisher information matrix, which is the inner product of the partial derivative of the strain with respect to the chirp mass:

\begin{align}\label{Eq27}
    \sigma^{-2}_{\mathcal{M}}&=\left(\frac{\partial h^*}{\partial\mathcal{M}}\Bigg|\frac{\partial h^*}{\partial\mathcal{M}}\right)\nonumber,\\
    &=2 \cdot 4\int^{{f}_{\text{max}}}_{{f}_{\text{min}}}{\rm d}{f}^\prime\frac{\left(\tilde{\frac{\partial h^*}{\partial\mathcal{M}}}({f}^\prime)\tilde{\frac{\partial h}{\partial\mathcal{M}}}({f}^\prime)\right)}{S_{\rm{t}}({f}^\prime)},
\end{align}

\noindent where $\tilde{h}$ is the Fourier transform of $h$. This relation is only valid for large SNR, which we liberally define as at least $8$, to drop $\mathcal{O}(\text{SNR}^{-2})$ terms. Calculating the partial derivatives using Eq.~\eqref{Eq5}, we obtain

\begin{align}\label{Eq28}
    {\frac{\partial h^*}{\partial\mathcal{M}}}&={\frac{\partial h}{\partial\mathcal{M}}}=\frac{5}{3}\frac{1}{\mathcal{M}_\text{tr}}h.
\end{align}

After taking the Fourier transform and using the relation $\tilde{h}^2=f^{-2}h_{\rm{c}}^2$. the inverse of the variance becomes

\begin{equation}\label{Eq29}
    \sigma^{-2}_{\mathcal{M}}=2 \cdot 4\int^{{f}_{\text{max}}}_{{f}_{\text{min}}}{\rm d}{f}^\prime\left(\frac{5}{3}\right)^2\frac{h_{\text{c}}^2({f}^\prime)}{S_{\rm{t}}({f}^\prime){f}^{\prime2}}\frac{1}{\mathcal{M}_\text{tr}^2}.
\end{equation}

In the above equation, we are using gas-free quantities due to our assumption that the vacuum model is the true model. The SNR-weighted apparent chirp mass between two frequencies is given by

\begin{equation}\label{Eq30}
    \mathcal{M}_\text{gas}=\sqrt{\frac{2 \cdot 4\int^{{f}_{\text{max}}}_{{f}_{\text{min}}}{\rm d}{f}^\prime\frac{h_{\text{c}}^2({f}^\prime)}{S_{\rm{t}}({f}^\prime){f}^{\prime2}}\mathcal{M}_{\text{gas}}^2({f^\prime})}{2 \cdot 4\int^{{f}_{\text{max}}}_{{f}_{\text{min}}}{\rm d}{f}^\prime\frac{h_{\text{c}}^2({f}^\prime)}{S_{\rm{t}}({f}^\prime){f}^{\prime2}}}}.
\end{equation}

Finally, the bias in the chirp mass estimation can be expressed as

\begin{equation}\label{Eq31}
    \Delta\mathcal{M}[\sigma]=\frac{\mathcal{M}_\text{gas}-\mathcal{M}_\text{tr}}{\sigma_{\mathcal{M}}},
\end{equation}

\noindent where $\mathcal{M}_\text{gas}$ and $\sigma_{\mathcal{M}}$ are calculated between observed frequencies $f_{\rm min}=f_{\rm i}$ and $f_{\rm max}=f_{\rm f}$. As customary, if $\mathcal{M}_\text{gas}$ is at least $2\sigma_{\mathcal{M}}$ away from $\mathcal{M}_\text{tr}$ then the gas induced bias is statistically significant. This criterion is equivalent to $\Delta\mathcal{M}[\sigma]\geq2$.

We find that $\Delta \mathcal{M}[\sigma]\lesssim10^{-4}$ for both set of binary/disc parameters in Fig.~\ref{Fig5} and Fig.~\ref{Fig6}. Therefore, we can infer that gas induces statistically insignificant bias when we recover the source chirp mass for an alpha disc and for an extrapolated super-Eddington disc (at $z=10$). The systematic error due to gas is negligible due to weak torques, and within the variance of the chirp mass measurement. However, if the source has higher SNR, the bias will become more significant. One can perform a more rigorous time-binning analysis to find signatures of gas in the parameter estimation \citep[see, e.g.][]{Gair2010}.

One could also analyze only the beginning of the signal, during which gas effects are most important as per Fig.~\ref{Fig7}. This analyses could occur if one performs parameter estimation `on the fly' during the event observation, which will happen during EM counterpart searches. Conversely, it could also be considered a way to search for gas signatures after the parameters are constrained post-merger. Hence, we consider the early inspiral signal until the SNR accumulates to 8, which occurs within the first $\sim3$ years of the event to enhance the gas-induced deviation in the GW waveform. We still find similar bias in chirp mass estimate as the entire signal because the relatively lower SNR of the early inspiral part also increases the measurement error, which compensates enhanced impact of the gas.

This section illustrates one does not need to consider the gas-induced bias in chirp mass estimate during parameter estimation because for all our system parameters, $\Delta\mathcal{M}[\sigma]\ll2$. 

\subsection{The hottest and densest discs}\label{Sec5.3}

In this section, we are interested in computing the minimum surface density $\Sigma_\text{Min}$ and the minimum speed of sound $c_\text{s,Min}$ (or, alternatively, minimum temperature\footnote{See Eq.~\eqref{EqA3} for the relation between the speed of sound and temperature.}) that would lead to an $\text{SNR}_{\delta\phi}\approx8$. In other words, we are looking for the critical gas values for distinguishing between purely GW-driven inspiral and gas-affected evolution. The reason for this exercise is to compute how strong the gas torque needs to be to make dephasing detectable. Additionally, since the phase shift scales with disc parameters via $\delta\phi\propto \Gamma_\text{vis}\propto\dot{M}\propto\alpha\Sigma\mach^{-2}$ then a detectable dephasing could be traced back to a constraint on the local gas quantities. One can also compare the limiting values of density and temperature to what is expected in galactic nuclei or the centres of gas-rich halos that may host these BH mergers.

\begin{figure}
\centering
\subfloat[]{\includegraphics[width=1\linewidth]{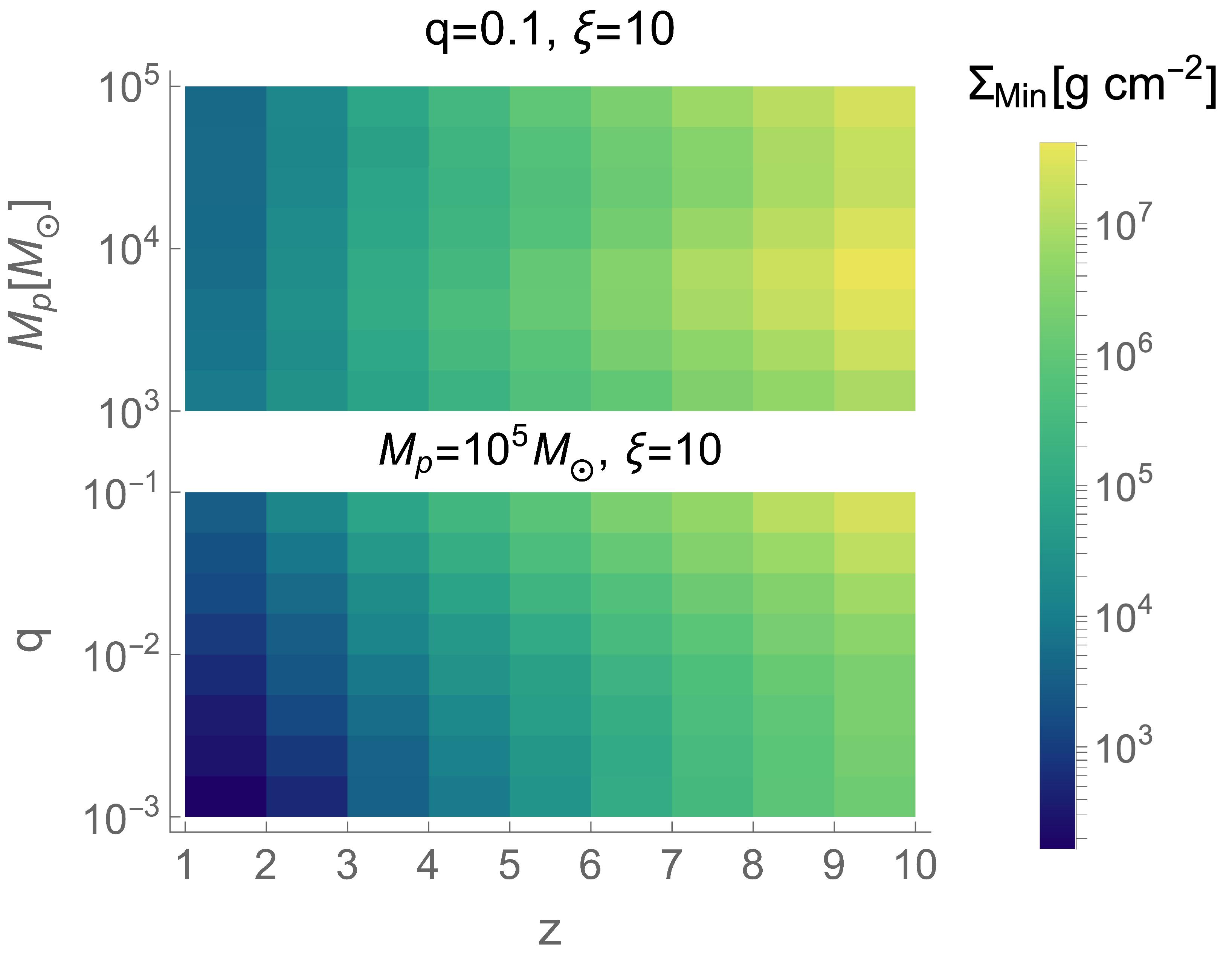}}\\
\subfloat[]{\includegraphics[width=1\linewidth]{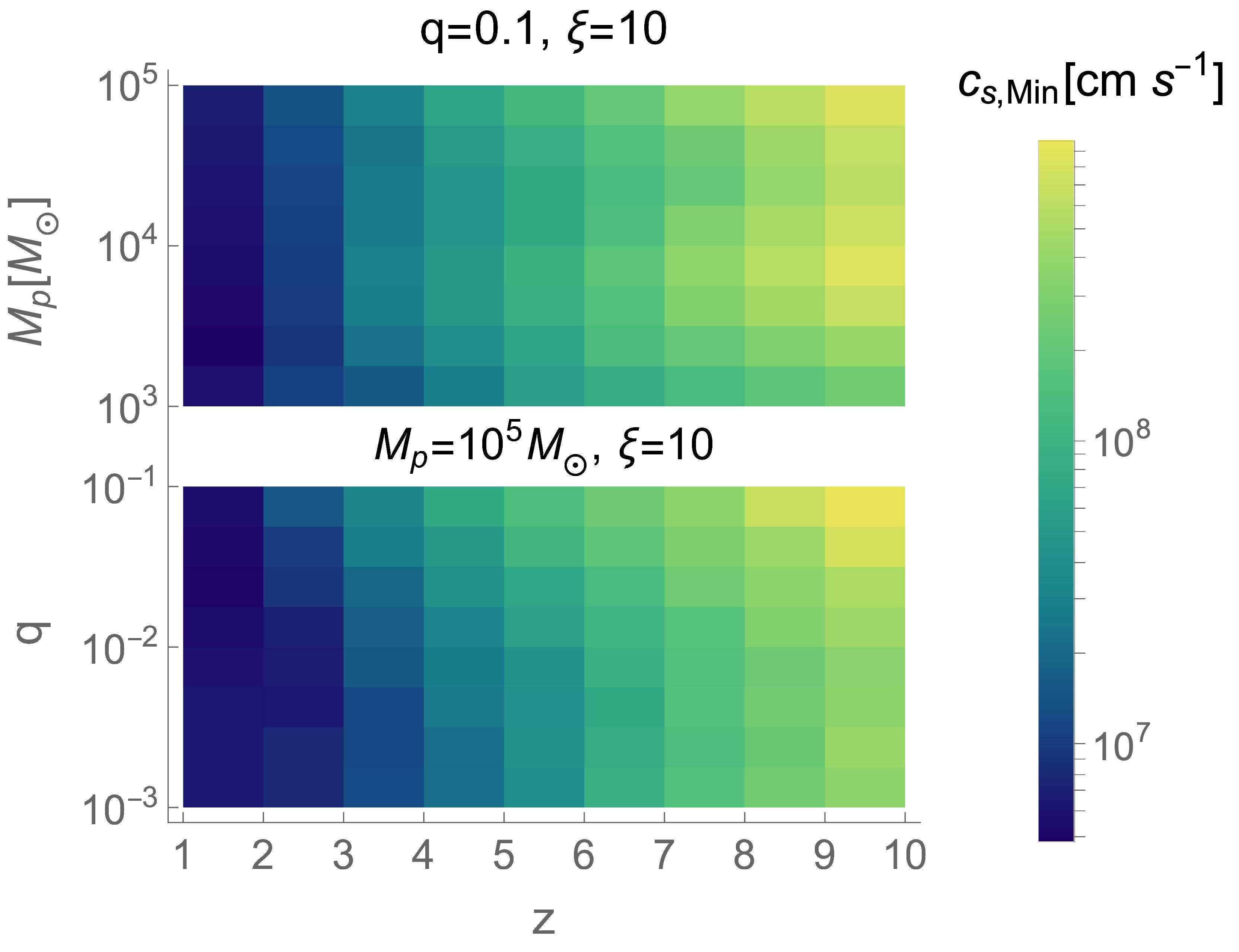}}
\caption{For the same binaries as in Fig.~\ref{Fig2}: (a) the minimum surface density and (b) the minimum speed of sound (when the binary enters the LISA window) of a thin accretion disc required to make the gas-induced dephasing detectable by LISA by having SNR$_{\delta\phi}\approx8$.}
\label{Fig8} 
\end{figure}

Again, we assume $\Gamma_\text{gas}=10\,\Gamma_\text{vis}$ to be the effective gas torque during the binary evolution. Because $\text{SNR}_{\delta\phi}\propto\delta\phi_\text{gas}\propto \Gamma_\text{gas}\propto\Sigma\mach^{-2}$, increasing the surface density or decreasing the Mach number would increase SNR$_{\delta\phi}$.\footnote{This is assuming that the radiative efficiency $\eta$ does not change, and that the gas torque always scales with $\dot{M}$.} In Fig.~\ref{Fig8}. we show $\Sigma_\text{Min}$ and $c_\text{s,Min}$ (when the binary enters the LISA window) for the same IMBHBs as in Fig.~\ref{Fig2}.

We can infer from this figure that, to detect phase shifts due to gas in binaries with redshifts $>7$, we need to have extremely high density $\Sigma\gtrsim10^7$~g~cm$^{-2}$ or speed of sound $c_{\rm{s}}\gtrsim3\times10^{8}$ cm s$^{-1}$. This translates into having accretion rates larger than $10\dot{M}_\text{Edd}$ for $\eta=0.1$. The higher speed of sound arises from our disc model assumption that a higher $\dot{M}$, which leads to stronger torques, requires higher disc temperature (or, lower Mach number).

\section{Discussion}\label{Sec6}

The detection of environmental signatures in GW waveforms from massive BHBs will be a unique opportunity to probe the surroundings of massive BHs, which we know very little about. Uncertainties remain in the disc's accretion flow, especially at super-Eddington rates, and the exact gas configuration and its properties are unknown in this regime. Inferring gas parameters -- mainly density and temperature -- from GWs could constrain accretion models at small scales (sub-pc separation). Most large-scale simulations of galactic nuclei and/or of galaxy formation in a cosmological context in the literature (see, e.g. \citealt{DeRosa2019}; \citealt{AmaroSeoane2022} and references therein) do not spatially resolve accretion on to the BHs. Instead, they assume some prescription within an unresolved region around the BH, which is usually larger than or equal to the BH's Bondi radius ($r_\text{Bondi}\equiv2GM_\text{BH}/c^2_{\rm{s}}$). Yet, the gas responsible for de-phasing in this work is at a smaller ($\sim100~r_{\rm{s}}\sim10^{-4}r_\text{Bondi}$) scale which is unresolvable in the EM domain. However, to constrain accretion flows at small scales using gas-induced dephasing in the GW waveform, we first require a model to connect phase shift to the accretion rate $\dot{M}$ and then a prescription to relate $\dot{M}$ to crucial disc parameters: density, temperature, and $\alpha$. In practice, we can only measure these parameters in a specific combination depending on the disc model, which leads to degeneracies. State-of-the-art MHD simulations of accretion discs around a $5\times10^8~\MSun$ SMBH suggest that inner disc densities are $\sim10^{4-6}$ g cm$^{-2}$ \citep{Jiang2019}. However, it is unclear if we can extrapolate this to discs around IMBHs. Present or future work that predicts gas densities near coalescence of BH seeds or active dwarf galaxies (e.g. some future high-resolution simulation) could help constrain gas imprints in GW detections by LISA. 

Conventionally, it is assumed that GWs heavily dominate the gas effects when binaries are in the late phase of the GW-drive inspiral (less than $\sim$100~$r_{\rm{s}}$). Indeed, the amount of gas contained within the binary orbit at those separations is several orders of magnitude lower than the binary mass (in our case, the enclosed gas mass is $\sim$10$^{-4}~\MSun$ within $100~r_{\rm{s}}$ for a $\sim$10$^5~\MSun$ binary). However, recent works have shown that gas torques may be stronger than previously anticipated \citep{MacFadyen2008,Haiman2009,Duffell2014,Farris2014,DOrazio2015,Kanagawa2018,Duffell2020,Munoz2020,Tiede2020,Derdzinski2019,Derdzinski2021}. \citet{Haiman2009} were amongst the first who argued that the gas torques can even become comparable to the GW torque in the LISA band for $\sim10^5~\MSun$ BHBs. Some of these works also predict both inward and outward migration, as well as fluctuating torques on massive BHBs, so that the overall outcome seems to be dependent on gas parameters and is not yet fully understood. Furthermore, more work is needed to model the late stage of inspiral in simulations that include both gas-driven torques and orbital energy dissipation by GWs \citep{Derdzinski2021}. Even if the orbital decay rate might be already controlled by GW emission in this relatively late stage of binary evolution, detecting deviations in a GW waveform due to gas would offer a unique opportunity to constrain  gas-driven torques at sub-pc scales. Our work is thus an attempt to explore these effects in a region of parameter space, that of IMBHBs, which, given their low masses, will spend more time in this late inspiral stage and have relatively high SNR.

Our results also have implications for IMBH seed growth models. For instance, $\sim$10$^5~\MSun$ IMBHs in the local Universe can either be the remnant of gas-deficient heavy seeds formed by direct collapse or the end point of repeated mergers and accretion phases of light seeds occurring in a gaseous environment \citep{Mezcua2019}. Therefore, the detection of gas-induced signatures in GWs from coalescing BHs at high redshift with masses consistent with light seeds could lend support to the latter seeding  model for IMBHs. Theoretical models of the growth of light seeds such as \citet{Sassano2021} predict mean BH accretion rates of $\sim$10$^{-2}$--$10~\MSun$~yr$^{-1}$ at $z\sim8$--10 and for a $10^5~\MSun$ BH, this is equivalent to $\sim$1--10$^4~\dot{M}_\text{Edd}$. This suggests that our scaling of $\dot{M}$ in Section~\ref{Sec5} is reasonable and that, in principle, one could extend that to even higher accretion rates. However, direct  radiation-MHD simulations of the accretion flow around an IMBH under the conditions of galactic nuclei at high redshift will be ultimately needed to calibrate phenomenological recipes for accretion in the super-Eddington regime.

The environment at high redshift (when galaxies are still in their early stages and are presumably more gas-dominated than at low redshift) does differ from that of the local Universe. \citet{Regan2009} show that, at $z\sim15$, it is difficult to have a gravitationally stable gas disc at the centre of a DM halo. Yet at $z\sim7$--11, there are a handful of observations and multiple candidates of full-fledged galaxies \citep[see, e.g.][]{Bouwens2015,Oesch2016,Oesch2018}, which, in principle, suggest the possibility of a steady-state AGN accretion disc configuration at their galactic nuclei as it occurs for galaxies at lower redshift. We also have several observations of AGN up to $z\sim7$ \citep{Padovani2017}. Indeed, according to \citet{MayerBonoli2019}, a merger of two gas-rich galaxies at $z>6$ leads to a formation of a gas accretion disc with gas surface density $\sim10^{5-6}$ g cm$^{-2}$ at the centre, which is a similar value as in our disc solutions (see Fig.~\ref{Fig3}). If halos hosting BHs merge, or if one halo forms multiple interacting BHs \citep{Regan2020}, the interaction between BHs and gas should be more dynamic. In this case, gas torques on a binary are uncertain but likely stronger than in the steady-state, sub-Eddington, low-mass discs we consider in this study. Therefore, at earlier stages of galaxy formation, our dephasing could be an underestimate. However, higher-strength torques could lead to weaker gas imprints due to faster BH binary coalescence.

Our results in Fig.~\ref{Fig8} show that if LISA is indeed able to detect dephasing in high-redshift IMBHBs, then we could in principle measure the BH accretion rate. Depending upon which birth environment we consider for a given BH seed could provide us with information on accretion and feedback scenarios. Also, the lack of a detected dephasing (Section~\ref{Sec5.1}) places constraints on the accretion rates and strength of the gas torque. If LISA detects an IMBHB and finds that there is a dephasing, then in principle, it can be associated with the gas torque the binary experiences during its evolution. Furthermore, if this gas torque scales with $\dot{M}$ (as it does in our model via the viscous torque), then the observed phase shift is a direct measurement of the accretion rate $\dot{M}$. However, if there is no dephasing, then either there is no gas or the gas torque (or $\dot{M}$) is weak. Moreover, if we detect an EM counterpart of this GW event, then this implies accretion on to the BHs. If there is no associated dephasing, then this is an upper constraint on the local gas properties under the assumptions of our model as per Fig.~\ref{Fig8}.

In the low-redshift Universe ($z\lesssim3$), there are now several candidates for accreting IMBHBs in the centres of dwarf galaxies (see, e.g. \citealt{Mezcua2017,Greene2020}). If any of these are in binaries that fall into the LISA band (at separation below $\lesssim100~r_{\rm{s}}$), gas signatures could be present in the waveform. However, constraining the accretion rates and understanding the gas disc parameters remains challenging. Some studies find observational signatures of radiatively inefficient accretion flows \citep{Wang2013,Mocibrodzka2014}, in which case the gas density and torques should be weak. Also, generally dwarf galaxies have low central densities, and indeed most of them do not have bulges (which suggests that it is difficult to form an IMBHB), although a few active ones appear to have a mass concentration at their centre. 

The gas torque expression we use in this study is $\Gamma_\text{gas}=\xi \Gamma_\text{vis}=\xi\dot{M}r^2\Omega$, which, for a constant accretion rate, depends upon the radial separation and orbital frequency as $\Gamma_\text{gas}\propto  r^{1/2}\propto f^{-1/3}$. Therefore, the dephasing $\delta\phi$, which depends upon $\dot{r}_\text{gas}\propto r^{1/2}\Gamma_\text{gas}$, has a frequency dependence of $\delta\phi\propto f^{-13/3}$. This frequency scaling is not degenerate with any standard PN order \citep{Blanchet1995,Will2004} and a high-SNR event should be able to distinguish it from PN corrections. However, attributing this dephasing to just the impact of gas torques is not possible unless we better understand different environmental effects on GWs, for example, the accretion on to the BHs \citep{Caputo2020}, effects of the thermal torque \citep{Hankla2020}, the presence of DM halos overdensities \citep{Kavanagh:2020cfn}, and possible third-body interactions \citep[see, e.g.][]{Samsing2018,Bonetti2018,Zwick2021,Rozner2022}, among others. Ideally, we should consider all of them in order to obtain a comprehensive picture, but we leave that to future work.

\subsection{Caveats}

A critical assumption for our detectability estimates in Section \ref{Sec5} is that we will have accurate enough GW waveforms for the source such that we can measure phase shifts of order radians. In reality, this will be challenging because when IMBHBs enter the LISA band, they will have non-negligible eccentricity and effective spin. Currently, vacuum waveform models do not span the whole range of possible spins and eccentricities. A tiny mismatch in the modelling of a waveform can induce dephasing of several orders of magnitude, making it difficult to constrain the actual gas-induced dephasing \citep[see, e.g.][]{Porter2010,Huwyler2012}. Therefore, our results underline the necessity of  developing accurate waveform catalogs in order to probe source environments.

Spin and eccentricity were both not considered in our work, while they could both play a role in the dynamics of the IMBHs during the early phase of the inspiral, in turn affecting the environmental signatures on the waveforms. The spin-orbit coupling occurs at 1.5~PN order \citep{Cutler1994}, whereas we assume a Newtonian treatment of the dynamics. Most AGN/SMBHs observed by X-ray reflection method are close to maximally spinning \citep{Reynolds2021}, which could be the case also for active dwarf galaxies. For maximally spinning BHs, the total orbital angular momentum ($=\mu\Omega r^2$) of a BH binary with $q\sim0.1$ (hence in the cavity regime) is about hundred times the absolute sum of individual spin angular momenta contributions at 1.5~PN order ($=(v_\phi/c)^3(G/c)[M^2_{\rm p}+(qM_{\rm p})^2]$) at $r_{\rm i}$, where gas effects are strongest in the LISA band. For $q\sim10^{-3}$, for which we expect a gap, the ratio of orbital angular momentum to the secondary BH spin angular momentum is close to a million. Therefore, naively we expect properly aligned spin(s) to affect the gas torques in the cavity regime more substantially than the Type-II regime. Moreover, a recent 3D general relativistic MHD simulation shows that the mini discs around BHs when the binary carves out a cavity are more massive around spinning BHs, suggesting that gas torques may change for  spinning BHs \citep{Combi2021}. A moderate eccentricity  could persist in the LISA band  (see, e.g. \citealt{Lima2020,Cardoso2021,DOrazio2021}), which could affect the gas flow around the IMBHBs, and therefore the magnitude and direction of torques they are subject to.

We only consider the quadrupole mode in this work. However, higher-order modes affect the GW emission for the non-spinning systems with mass ratios $\gtrsim0.25$ \citep[see, e.g.][]{Pekowsky2013, Capano2014}. Therefore, including higher-order modes is an essential next step to get more accurate results.

Another critical assumption in this work is the linear addition of the orbital evolution rate due to GWs and gas ($\dot{r}=\dot{r}_\text{\scriptsize{GW}}+\dot{r}_\text{gas}$). While this should be reasonable for low mass ratios ($q\sim10^{-3}$; as suggested by \citealt{Derdzinski2021}), for which the gas torques do not evolve significantly due to the GW-driven inspiral, it is less certain for higher mass ratios. \citet{Tang2018} suggest that gas morphology changes drastically as BHBs approach coalescence. Therefore, high-resolution numerical simulations are needed to check the validity of the linear addition of orbital evolution rates. Morover, if this assumption breaks down, then this will further affect phase accumulation and parameter estimation.

Even in the simple case of non-spinning IMBHBs on circular orbits treated here, deviations from standard linear torque theory may arise, such as torque variability on sub-orbital time-scales or sign changes (as shown in \citealt{Derdzinski2019,Derdzinski2021}). These will lead to more complex yet informative signatures \citep{Zwick2022},  whose detectability in a population of IMBHBs with varying properties and redshift will be studied in future work.

In Fig.~\ref{Fig6}, we consider super-Eddington accretion rates. The critical assumption is that an IMBHB evolution event inside a LISA four-year observation window coincides with a super-Eddington accretion episode. While there is no work which address this directly, \citet{Massonneau2022} show that the duty cycle\footnote{The fraction of the time BH accretes at super-Eddington rates in their simulation over $282$ million years.} of such an episode for a central BH could be as small as $\sim0.01$ if the AGN feedback for higher accretion consists of both jet-like outflows and thermal energy release, or it could be as high as $\sim0.1$ if the AGN feedback only releases thermal energy. This makes the detection of dephasing by LISA in high-redshift IMBHBs highly unlikely, \emph{unless} the accretion episode is triggered due to a merger of galaxies, in which case the BHB formation can occur in the presence of a large amount of gas. However, in the case of dwarf galaxies, AGN activity due to galaxy mergers seems to be not as correlated as for the massive galaxies \citep{Stierwalt2021}.

\subsection{Comparison to previous work}

Most of the work on the dephasing in the LISA band due to environmental effects is focused on EMRIs rather than near-equal mass ratio BHBs. The primary motivation for this was that EMRIs spend many more orbits away from the merger in the LISA band, where it does not significantly chirp, and GWs do not entirely overwhelm non-vacuum effects, allowing the gas to impart detectable dephasing in the waveform. Also, only recently we have found strong evidence for the existence of IMBHs, which suggests a more likely possibility of IMBHBs. This work has shown that, if they exist, IMBHBs have a similarly high number of cycles as EMRIs and enough SNR to observe dephasing in the LISA band.

The earlier work on dephasing is by \citet{Yunes2011} and \citet{Kocsis2011}, who consider EMRIs and study them assuming only a one year LISA observation window with different final separations. We computed dephasing assuming their LISA and binary parameters (their model II$\alpha$) using our setup in this work and found the same order-of-magnitude dephasing for the merger in the LISA band. Recent studies by \citet{Derdzinski2019} and \citet{Derdzinski2021} have similar LISA specifications as in this work, and their gas torque is close to $0.1\Gamma_{\rm vis}$. They also obtain detectable dephasing only if the surface density is larger than $\sim10^4$ g cm$^{-2}$ for a $10^{-3}$ mass ratio binary.

Comprehensive work by \citet{Barausse2014} is devoted to BHB evolution in a thin disc and the impact of self-gravity, migration, and dynamical friction on the GW waveform. While mainly focusing on EMRIs, they argue that their results can be extrapolated to near-equal massive BHBs to obtain order-of-magnitude estimates. The effect of self-gravity is not relevant in our case, as BHBs are much more massive than the enclosed gas mass. In the migration case, they consider both Type-I and Type-II migration only for a $\beta$ thin disc model \citep{Sakimoto1981}, which has a higher surface density than the $\alpha$ disc, and showed that while the gas torques could be important with respect to GW torques, massive binaries will destroy the disc, hence it is not relevant. This is based on a classical argument that the gas does not follow the binary till the merger, which has been recently shown not to be the case in numerous CBD simulations discussed before. Furthermore, while they did mention the possibility of a CBD formation around a massive BH, they conclude that the Type-II torque is only important outside the LISA band. However, their detectability threshold (i.e. gas effects are only relevant if the gas-to-GW torque ratio $\Gamma_{\rm gas}/\Gamma_{\rm \scriptsize{GW}}$ is unity) sets the bar too high as the impact of weaker gas torques ($\Gamma_{\rm gas}/\Gamma_{\rm \scriptsize{GW}}\ll1$) can also become significant if added over many cycles in the LISA band as shown in this work. For the  regime of dynamical friction, they compute phase shifts for BHBs, which are fully embedded in a thin disc, in their last year before reaching ISCO in the LISA band. For our fiducial parameters, dynamical friction suggests a phase shift of $\sim50000$ radians using their methodology, which is much higher than what we show in Fig.~\ref{Fig5}. However, we caution that the dynamical friction estimate is not fully applicable in this case. In fact, the gas mass within the orbit of the BHBs at sub-pc scale separations is always much smaller than any of the two BH masses, so that the disturbance that the secondary BH induces in the background cannot be treated as a small density perturbation as assumed in Chandrasekhar's theory \citep{Chandrasekhar1942}. Also, our BHB is not fully submerged in the disc but instead creates a gap/cavity and leads to a much weaker gas imprint.

\section{Conclusions}\label{Sec7}

In this paper, we study gas effects on IMBHBs (where both companions have masses in the range of $10^2$--$10^5$~M$_{\sun}$) via GWs. Our fiducial binary consists of two non-spinning IMBHs in a circular orbit. We analyse this binary from the moment it enters the LISA frequency band, which occurs at a rest-frame binary separation of around $100$ times the Schwarzschild radius of the primary BH, to when it exits the LISA band or the binary separation reaches the ISCO of the primary BH. IMBHBs at redshifts 1--10 with mass ratios $10^{-3}$--$10^{-1}$ spend $\sim$10$^3$--$10^5$ orbits with SNR~$\sim$~5--350 in the LISA band (see Fig.~\ref{Fig2}). Therefore, if they exist, most of the IMBHBs in this parameter space will be detectable by LISA (assuming a detectability threshold of SNR $\geq8$). We embed this binary at the centre of a thin gas disc (see Appendix~\ref{AppA} for details) with a viscosity coefficient $\alpha=0.01$ and an accretion rate $\dot{M}=0.1\dot{M}_\text{Edd}$. For central BH masses of $10^3$--$10^5$~M$_{\sun}$, fiducial values of the Mach number $\mach$ and surface density $\Sigma$ of the disc are $80$ and $2\times10^5$ g cm$^{-2}$, respectively (see Fig.~\ref{Fig3}). We consider various masses, mass ratios, and redshifts relevant for IMBHBs. We itemize our findings below.

\begin{itemize}

    \item IMBHBs embedded in a thin gas disc are expected to be in a Type-II/cavity migration regime due to secondary BH/binary opening a gap/cavity during its evolution in the LISA band (see Section~\ref{Sec4} and Appendix~\ref{AppB}).
    
    \item In the Type-II/cavity regime, we find the gas torque scales with the viscous torque and the exact scaling $0.1\lesssim\xi\lesssim10$ depends non-linearly on the disc and binary parameters (see Section~\ref{Sec4}).
    
    \item Gas-induced dephasing in the GW waveforms is detectable by LISA for a subset of IMBHBs. For $\xi=0.1$, up to $z \sim 3$ for $q\lesssim0.05$. For $\xi=10$, until $z \sim 7$ for low mass ratios $q\approx10^{-3}$ or up to $z \sim 5$ for $q\approx0.1$ (see Fig.~\ref{Fig5}). The maximum observable phase shift is $\sim$96~radians, or $\sim$15~GW cycles over an observation of $\sim$80000~cycles.

    \item Scaling our disc solutions to super-Eddington accretion rates will make phase shifts for high-redshift binaries detectable (see Fig.~\ref{Fig6}). While our disc solutions are no longer applicable, our inference should give an order-of-magnitude estimate.

    \item Using the simplified Fisher information matrix analysis, we find that gas signatures on the recovered chirp mass of our fiducial binary induce insignificant statistical bias irrespective of the accretion rate (see Section.~\ref{Sec5.2}).

    \item If LISA indeed detects dephasing due to gas in GW waveforms of high-redshift IMBHBs, then this places a lower constraint on the disc surface density $\Sigma\gtrsim10^7$~g~cm$^{-2}$ or the speed of sound $c_{\rm{s}}\gtrsim3\times10^8$ cm s$^{-1}$ under our model's assumptions (see Fig.~\ref{Fig8}). 

\end{itemize}

\section{Data availability statement}
The data underlying this article will be shared on reasonable request to the authors.

\section*{Acknowledgements}
The authors acknowledge support from the Swiss National Science Foundation under the grant 200020\_192092. AD also acknowledges support from the Tomalla Foundation for Gravity Research. We acknowledge Fr\'{e}d\'{e}ric Masset, Mar Mezcua, John Regan, and Shubhanshu Tiwari for insightful discussions. We also thank Zolt\'an Haiman, Nicholas Stone, and Alejandro Torres-Orjuela for useful comments on the manuscript. We also acknowledge use of the Mathematica software \citep{Mathematica} and NumPy \citep{harris2020array}.  

\scalefont{0.94}
\setlength{\bibhang}{1.6em}
\setlength\labelwidth{0.0em}
\bibliographystyle{mnras}
\bibliography{imbh}
\normalsize

\appendix
\section{Accretion disc solution}\label{AppA}

We adopt a geometrically thin, steady-state, sub-Eddington, Keplerian accretion disc model around a single IMBH. 
The model follows the inner disc equations in \citealt{Derdzinski2022}, similar to the model in \citet{Sirko2003}. It is essentially a solution of the \citet{ShakuraSunyaev1973} thin disc equations that includes gas and radiation pressure. As in \citet{ShakuraSunyaev1973}, we adopt the turbulence-motivated viscosity prescription, where the kinematic viscosity $\nu = \alpha c_{\rm{s}}^2 \Omega^{-1}$ is parametrized by the coefficient $\alpha<1$.

The effective temperature at the disc surface arises from internal heating via viscous dissipation and is related to the constant accretion rate $\dot{M}$ via
\be\label{EqA1}
\sigma T_{\rm eff}^4 = \frac{3}{8 \pi} \Omega^2 \dot{M'},
\ee
where $\sigma$ is the Stefan-Boltzmann constant and $\dot{M'} = \dot{M}(1-(r_{\rm min}/r)^{1/2})$, which arises from applying a 
zero-torque boundary condition at the disc inner radius $r_{\rm min}$ (this ensures that the viscous torque goes to zero at $r_{\rm min}$). Assuming the disc is in a steady state, the surface density is dependent on the rate of viscous inflow:
\be\label{EqA2}
\Sigma = \frac{\dot{M'}}{3 \pi \nu}.
\ee

The speed of sound depends on both gas and radiation pressure:
\be\label{EqA3}
c_{\rm{s}}^2 = \frac{p_{\rm gas}+p_{\rm rad}}{\rho} = \frac{k_{\rm B}}{\mu m_{\rm H}} T_{\rm mid} + \frac{1}{3} a_{\rm rad} \frac{T_{\rm mid}^4}{\rho},
\ee
where $m_{\rm{H}}$ is the mass of hydrogen, $k_{\rm B}$ is the Boltzmann constant, $a_{\rm rad}$ is the radiation density constant, and we take the mean molecular weight $\mu$ for an ionized gas to be $0.62$.

Assuming that energy transport from the disc midplane to the surface is radiative, the midplane temperature is
\be
T_{\rm mid}^4 = \left(\frac{3}{8}\tau_{\rm opt} + \frac{1}{2} + \frac{1}{4 \tau_{\rm opt}}\right) T_{\rm eff}^4,
\label{eq:diffusion}
\ee
where $\tau_{\rm opt}=({1}/{2}) \kappa \Sigma$ is the optical depth. The Rosseland mean opacity $\kappa$ is computed by a piecewise function following the form
\be
\label{eq:opacity}
\kappa = \kappa_0 \rho^a T^b
\ee
where $\kappa_0$, $a$, and $b$ are defined over regimes of $\rho$ and $T_{\rm mid}$ following the fits from \citet{Bell1994}. In the regions of interest for IMBHs, the relevant sources of opacity are electron scattering and bound-free + free-free transitions. 

The scale height $h$ depends on the speed of sound by
\be\label{EqA4}
h = \frac{c_{\rm{s}}}{\Omega},
\ee
which also determines the structure and volume density at each radius by
\be\label{EqA5}
\rho= \frac{\Sigma}{2 h}.
\ee

This gives us a system of five equations with five variables: 
$\rho(r)$, $\Sigma(r)$, $T_{\rm mid}(r)$, $h(r)$, and $c_{\rm{s}}(r)$, which can be solved numerically at each radius $r$ given a choice of central mass $M_{\rm p}$, accretion rate $\dot{M}$, and viscosity coefficient $\alpha$. 

The disc can be described by a azimuthal Mach number $\mach = v_{\phi}/c_{\rm s}$, where $v_{\phi}$ is the azimuthal velocity. This can also be written in terms of the disc scale height: $\mach = r/h$. For simplicity, we describe our disc parameters with the surface density $\Sigma$ and the corresponding Mach number $\mach$, which are both used typically in migration torque expressions. 
Radial profiles of these quantities are shown in Fig.~\ref{Fig3} up to 1000 $r_{\rm s}$ for $\alpha=0.01$, $\dot{M}=0.1\dot{M}_\text{Edd}$, and central masses $M~=~10^3,~10^4,$ and $10^5$~M$_{\sun}$. Note that the gas temperature implied by the accretion rate is primarily in the regime where bound-free and free-free scattering dominate the opacity. For the highest IMBH mass, the inner region of the disc transitions to electron scattering (the transition is sharp due to the adoption of piecewise opacity laws). 

These profiles are used to inform the torque strengths and gap/cavity opening estimates in Figs.~\ref{Fig4}. Given that the radial and BH mass dependence of $\Sigma$ and $\mach$ is relatively weak, in calculations where the system parameters are interpolated across IMBH masses, we use the constant normalizations defined in Eq.~\ref{Eq9}.
\section{Type-I or Type-II migration?}\label{AppB}
In this section, we are interested in finding the mass ratio at which a given binary transitions from the Type-I to Type-II migration regime. Therefore, we study three conditions discussed in the literature to determine when to expect the secondary BH to open up an annular gap in a gas disc and whether it will influence the binary evolution. 

First, we consider the standard \citet{Crida2006} criterion for gap opening\footnote{This criterion has been further verified by \citet{Duffell2013} for our choice of $\alpha=0.01$.}. This arises from comparing  gravity, viscosity, and pressure, and expresses the fact that the gap is opened when the gravitational torque exerted by the smaller companion on the surrounding disc mass, which tends to push matter to larger distances, cannot be compensated by pressure and/or viscous forces. It can be formulated as
\begin{equation}\label{EqB1}
   \frac{3}{4}\frac{h}{r}\left(\frac{3}{q}\right)^{\frac13}+\frac{50\nu}{q\Omega r^2}\lesssim1\implies\frac{3}{4}\frac{1}{\mach}\left(\frac{3}{q}\right)^{\frac13}+\frac{50\alpha}{q\mach^2}\lesssim1.
\end{equation}

For our fiducial parameters $\alpha=0.01$ and $\mach=80$, above criterion is satisfied for any $q\gtrsim10^{-4}$. Therefore, for our system parameters, we always expect a gap. However, even if a gap is initially opened it may not be maintained during migration or not sustained against viscous refilling.  Hence, we need to consider different time-scales. 

The second condition expresses that it is still possible for the secondary BH to escape the gap. Therefore, for the BH to not cross its Horseshoe radius $R_\text{HS}\equiv 2.4 r(q/3)^{1/3}$ (defined as the gap width following \citealt{Paardekooper2008,Malik2015,Muller2018}), its crossing time-scale $t_\text{cross}\equiv R_\text{HS}/|\dot{r}|$ (which represents the time taken by the secondary companion to migrate $R_\text{HS}$) should be more than the gap-opening time-scale $t_\text{gap}$ \citep{Hourigan1984}. We define $t_\text{gap}$ as 10 times half of the libration time\footnote{Libration time is the time taken by a perturber to remove 10~per cent of the disc mass inside its horseshoe radius $R_\text{HS}$.} \citep{Masset2008}. Therefore, the estimate of the gap-opening time-scale is

\begin{equation}\label{EqB2}
    t_\text{gap}=10\times\frac{2}{3}\frac{r}{R_\text{HS}}t_{\rm{0}}\implies  t_\text{gap}\approx8.6\left(\frac{q}{0.1}\right)^{-\frac13}t_{\rm{0}},
\end{equation}
\noindent where $t_{\rm{0}}\equiv2\pi/\Omega$ is the the orbital time period. It is important to note that this time-scale only indicates the time taken to clear out the gap, which could be much shorter than the time required to reach a steady-state gap formation. However, for our fast in-spiraling GW-driven binary, we can take this as a relevant gap-opening  time-scale. Also, for a GW-dominated binary, we can approximately take $\dot{r}=\dot{r}_\text{\scriptsize{GW}}$ even in the presence of gas due to weak gas torques as per Fig.~\ref{Fig4}. This condition can be expressed as

\begin{subequations}
\begin{align}
    &~ t_\text{gap}<t_\text{cross},\label{EqB3a}\\
    \implies&1.59\times10^{-4}\sqrt{1+q}\left(\frac{q}{0.1}\right)^{\frac13}\left(\frac{r}{100r_{\rm{s}}[M_{\rm{p}}]}\right)^{-\frac52}<1\label{EqB3b}.
\end{align}
\end{subequations} 
The above criterion is satisfied for any mass ratio $q$ we consider in this work for separations $r\gtrsim5r_{\rm{s}}[M_{\rm{p}}]$.

The third condition states that, even if a gap can be opened according to equation \eqref{EqB1}, it can be refilled by viscous diffusion. For it to not refill the gap, $t_\text{gap}$ should be smaller than the viscous diffusion time-scale $t_\text{vis}\equiv R_\text{HS}^2/\nu$ which quantifies the time taken to smooth out density gradients for a gap of width $R_\text{HS}$. Therefore, this condition is

\begin{subequations}
\begin{align}
    &~ t_\text{gap}<t_\text{vis},\label{EqB4a}\\
    \implies&1.42\times10^{-4}\left(\frac{q}{0.1}\right)^{-1}\left(\frac{\mach}{80}\right)^{-2}\left(\frac{\alpha}{0.01}\right)<1\label{EqB4b}.
\end{align}
\end{subequations}

This criterion is sensitive to both mass ratio and disc parameters. For $\alpha=0.01$ and $\mach=80$, it is satisfied for any $q\gtrsim10^{-5}$. This is in agreement with \citet{Duffell2013}, who show that even a small secondary BH can open a gap given suitable disc parameters. But for a higher viscosity coefficient or a lower Mach number, Type-I to Type-II transition happens for relatively higher mass ratios.

For our systems of interest both Eqs \eqref{EqB3b} and \eqref{EqB4b} are always satisfied. Therefore, we do expect a gap to open, be maintained during the secondary BH migration and sustained against viscous refilling. Hence, binaries are in the Type-II regime for $q<0.04$ or, in the cavity regime for higher mass ratios. 

\bsp %typesetting comment
\label{lastpage}
\end{document}